\documentstyle{article}
\newcommand{\lr}{LRSM$\,\,$}

\newcommand{\p}{\partial}
\newcommand{\nonum}{\nonumber\\}
\newcommand{\be}{\begin{equation}}
\newcommand{\ee}{\end{equation}}
\newcommand{\ba}{\begin{eqnarray}}
\newcommand{\ea}{\end{eqnarray}}
\newcommand{\bd}{\begin{displaymath}}
\newcommand{\ed}{\end{displaymath}}

\newcommand{\ma}[1]{_{\mathop{}_{#1}}}
\renewcommand{\a}{&}
\makeatletter
 
          \@addtoreset{equation}{section}
       \makeatother
\begin{document}
\title{ \bf
Left-right symmetric gauge model \\
in a noncommutative  geometry on 
${ M_{ 4}}\times { Z_4}$ 
}
\author{Yoshitaka  {\sc Okumura}\thanks{
e-mail address: okum@isc.chubu.ac.jp}\\
{\it Department of Natural Science, Chubu University, Kasugai 487, Japan}
}
\date{}
\maketitle
\thispagestyle{empty}
\begin{abstract}
The left-right symmetric gauge model with the symmetry of 
$SU(3)_c\times SU(2)_{_L}\times SU(2)_{_R}\times U(1)$
is reconstructed in a new scheme
of the noncommutative differential geometry (NCG)
on the discrete space $M_4\times Z_4$ which is a product space of
Minkowski space and four points space.
The characteristic point of this new scheme is to take the fermion
field to be a vector in a 24--dimensional space which contains all
leptons and quarks. Corresponding to this specification, all gauge and
Higgs boson fields are represented in $24\times 24$ matrix forms.
We incorporate two Higgs doublet bosons $h$ and $SU(2)_{_R}$
adjoint Higgs $\xi_{_R}$ 
which are as usual transformed as $(2,2^\ast,0)$
and $(1,3,-2)$ under $SU(2)_{_{L}}\times SU(2)_{_{R}}
\times U(1)$, respectively. 
Owing to the revise of the algebraic rules in a new NCG, 
we can obtain the necessary potential and interacting terms between these 
Higgs bosons which are responsible for giving  masses to the particles 
included. Among the Higgs doublet bosons, one CP-even scalar boson
survives in the weak energy scale and other four scalar bosons acquire
the mass of the $SU(2)_{_{R}}\times U(1)$ breaking scale,
which is similar to the situation in the standard model.
  $\xi_{_R}$ is responsible 
to spontaneously break $SU(2)\ma{R} \times U(1)$ down to $U(1)\ma{Y}$
and so well explains the seesaw mechanism. 
Up and down quarks have the different masses through the vacuum expectation 
value of $h$.
\vskip 0.5cm
\noindent
PACS number(s):11.15.Ex,02.40.Hw,12.15-y
\end{abstract}
\section{Introduction}
The efforts made so far to understand 
the Higgs mechanism responsible for the spontaneous breakdown of the
gauge symmetry have proposed many attempts such as technicolor model,
Kaluza-Klein model, strong coupling model and the approach based
on the noncommutative differential geometry(NCG) on the discrete space,
among which the NCG approach does not demand any extra physical modes
except for the usual modes, contrary to others and provides
the unified picture of the gauge  and Higgs boson fields 
as the generalized 
connection in NCG on the discrete space $M_4\times Z_{_{N}}$.
\par
Since Connes \cite{Con} has proposed 
the first idea to understand the Higgs mechanism
in NCG on the discrete space $M_4\times Z_2$,
 many works have been done so far 
${\cite{Cham}\sim\cite{WS}}$. 
The present author \cite{SU5}, \cite{NCGNP} 
has also developed an unique version of NCG approach
which is a natural extention of the differential geometry 
on the ordinary continuous space.
We define the generalized gauge field on the principal bundle with the 
base  space $M_4\times Z_{_N}$ by adopting one-form bases $dx^\mu$
in $M_4$ and $\chi_k(k=1\cdots N)$ in $Z_{_N}$. The generalized 
gauge field introduced in such a way contains both the ordinary gauge field 
and the Higgs field, which realizes the unified picture 
of the gauge and Higgs fields on the same footing.
According to our formulation of NCG, we have constructed 
the standard model \cite{FSTM},
the left-right symmetric gauge model(LRSM) \cite{WSLR}, 
SU(5) and SO(10) grand unified theories \cite{SU5},\cite{SO10}.
Generally speaking, the NCG approach imposes the severe constraints on 
the Higgs potential and interacting terms.
However, we could 
succeed in reconstructing the standard model \cite{FSTM}
because there exists one kind of the Higgs boson in this model.
We could perform the renormalization group analysis for the Higgs boson mass
with the initial condition $m_{_H}=\sqrt{2}m_{_W}$ strongly suggested 
to hold at the energy of the unification of electromagnetic and weak
interactions ($g=\sqrt{5/3}g'$ for the coupling constants of
$SU(2)_{_L}$ and $U(1)_{_Y}$ gauge group). The Higgs boson mass
is predicted to be 164GeV for the top quark mass 175GeV \cite{RGNCG}. 
On the other hand, we were not so successful to reconstruct the left-right 
symmetric gauge model \cite{WSLR} because we could not introduce the
sufficient Higgs potential and interacting terms owing to the severe
constraints by the algebraic rules in our previous NCG, 
 though the Higgs kinetic terms
are nicely introduced as curvatures on the discrete space. 
However, 
\lr is still alive as a model with the intermediate symmetry of the 
spontaneously broken $SO(10)$ grand unified theory(GUT) and are
  very important to contain the two-doublet fields $\phi_1$, $\phi_2$ 
and the $SU(2)_{_R}$ adjoint filed $\xi_{_R}$. These Higgs bosons
play important roles in giving  masses to up and down quarks and
inducing the seesaw mechanism 
for the left and right handed neutrinos \cite{Gell}.
In order to improve our previous reconstruction of \lr \cite{WSLR}
we relax in this article the algebraic rules to 
obtain the necessary terms for the Higgs potentials and interactions.
\par
This paper consists of five sections. The next section presents 
our noncommutative differential geometry on the discrete space
$M_4\times Z_{\mathop{}_{N}}$. 
The generalized gauge field is defined there 
and a geometrical
picture for the unification of the gauge and Higgs fields is realized. 
An algebraic rule will be modified so as to present the Higgs potential
terms sufficient to explain the phenomenology of \lr.
The third section  provides the reconstruction 
of the Yang-Mills-Higgs sector in \lr. 
The characteristic point  is to take the fermion
field to be a vector in a 24--dimensional space which has the
same placement of leptons and quarks as in $SO(10)$ GUT.
Corresponding to this specification, all gauge and
Higgs boson fields are represented in $24\times 24$ matrix forms.
As the Higgs bosons, we incorporate 
$h$ and $\xi\ma{R}$ transformed as $(2,2^\ast,0)$ and
$(1,3,-2)$ for $SU(2)\ma{L}\times SU(2)\ma{R}\times U(1)$, respectively.
After the construction of Yang-Mills-Higgs sector, the gauge and Higgs
boson masses are investigated. In section 4,
the fermionic sector of \lr is proposed after the 
presentation of  the general formulation.
It will be shown that 
the fermionic Lagrangian including lepton and quark sectors together 
is expressed in very compact form by use of 
the $24-$dimensional left and right fermion fields.
Lagrangian for the leptonic sector is nicely constructed 
so that the seesaw mechanism \cite{Gell} works well.
Quark sector contains the generation mixing and up and down quarks acquire
the masses through the vacuum expectation value of $h$.
 The last section is devoted to concluding remarks.
\noindent
\section{ Noncommutative  geometry on the discrete space 
                              $M_4\times Z_{\mathop{}_{N}}$ }
The NCG on the discrete space $M_4\times Z_{_N}$
has been formulated in Ref.\cite{NCGNP}, which we here briefly review
by fitting it to the presentation in this article.
We first define the generalized gauge field on the principal bundle 
with the base space $M_4\times Z_{_N}$ and the structure group of
the corresponding internal symmetry. 
Let us denote  the generalized gauge field 
${\cal A}(x,n)$ which is
represented in the  differential form of degree one as,
\be
      {\cal A}(x,n)=\sum_{i}a^\dagger_{i}(x,n){\bf d}a_i(x,n) 
          + \sum_j b_j^\dagger(x,n){\bf d}b_j(x,n), \label{2.1}
\ee
where the $a_i(x,n)$ and $b_j(x,n)$ are the square-matrix-valued functions
relating to the internal symmetry and are assumed to be 
$[a_i(x,n), b^j(x,n)]=0$.
The subscript $i$ and $j$ are   variables of the extra
internal space which we cannot now identify.  
Now, we simply regard $a_i(x,n)$  as a more fundamental
field  to construct gauge  and Higgs fields. 
The reason why $b_j(x,n)$ is incorporated is that the color gauge field
$G_\mu(x)$ is composed of $b_j(x)$, whereas the gauge fields of
the flavor symmetry and the Higgs field are composed of $a_i(x,n)$,
as shown later. 
However, in order to retain the consistent definition of gauge field,
the following condition has to be imposed:
\ba
&&    \sum_{i}a_{i}^\dagger(x,n)a_{i}(x,n)= 1, 
  \label{2.1a}\\
&&   \sum_{j}b_{j}^\dagger(x,n)b_{j}(x,n)= \frac1{g_s} \label{2.1b}
\ea
with the constant $g_s$ relating to the coupling constant of the
color gauge field.
${\bf d}$ in Eq.(\ref{2.1}) is the generalized exterior 
derivative defined as follows:
\ba 
      && {\bf d}=d + d_\chi=d+\sum_k d_{\chi_k}, \label{2.2}\\  
      && da_i(x,n) = \p_\mu a_i(x,n)dx^\mu, \label{2.3}\\
      && db_j(x,n) = \p_\mu b_j(x,n)dx^\mu, \label{2.4}\\
      && d_{\chi_k} a_i(x,n)=[-a_i(x,n)M_{nk}\chi_k 
                        + M_{nk}\chi_k a_i(x,n)], \label{2.5}\\
      && d_{\chi_k} b_j(x,n)=0 \label{2.6}
\ea
where $dx^\mu$ and  $\chi_k$ are the 
 one-form bases, taken to be dimensionless, in Minkowski space 
$M_4$ and in the discrete space $Z_{_N}$, respectively. 
We have introduced the $x$-independent matrix $M_{nm}$ 
generally expressed in the rectangular matrix
whose hermitian conjugation is given by $M_{nk}^\dagger=M_{kn}$. 
The matrix $M_{nk}$ turns out to determine the scale and pattern of 
the spontaneous breakdown of the gauge symmetry. Thus, the symmetry
breaking mechanism is encoded in the $d_\chi$ operation.
We determine  several important algebraic rules
in our NCG. Detailed descriptions of our NCG are shown in Ref.\cite{NCGNP} and
we here exhibit it concisely.
Let us at first introduce the shifting rule 
described by
\be
            \chi_k a_i(x,n) = a_i(x,k)\chi_k,  \label{2.7}
\ee
which  shows how to change the discrete variable 
when $\chi_k$ is shifted from left to right in an equation and
makes the product of matrix valued functions consistently calculable.
Inserting Eqs.(\ref{2.2})$\sim$(\ref{2.6}) into Eq.(\ref{2.1})
and using Eq.(\ref{2.7}),
${\cal A}(x,n)$ is rewritten as
\be
 {\cal A}(x,n)=A(x,n) + G(x,n) \label{2.8}
\ee
with the definitions that
\ba
   &&  A(x,n)={A}_\mu(x,n)dx^\mu
               +\sum_k{\mit{\mit\Phi}}_{nk}(x)\chi_k, \label{2.8a}\\
   &&  G(x,n)= G_\mu(x,n)dx^\mu, \label{2.8b}
\ea
where
\ba
&& A_\mu(x,n) = \sum_{i}a_{i}^\dagger(x,n)\p_\mu a_{i}(x,n), \label{2.9}\\
&&     {\mit{\mit\Phi}}_{nk}(x) = \sum_{i}a_{i}^\dagger(x,n)
         \,[-a_i(x,n)M_{nk} 
            + M_{nk}a_i(x,k)]. \label{2.10}\\
&&   G_\mu(x,n)=\sum_jb_j^\dagger(x,n)\p_\mu b_j(x,n).\label{2.11}
\ea
Here, $A_\mu(x,n)$, ${\mit{\mit\Phi}}_{nk}(x)$ and $G_\mu(x,n)$
 are identified with the gauge field of the flavor symmetry, 
 the Higgs field and the color gauge field,  respectively. 

 \par
We define the generalized field strength as
\be
  {\cal F}(x,n)= F(x,n)+ {\cal G}(x,n), \label{2.13}
\ee
where
\ba
 \a F(x,n)= {\bf d} A(x,n) +{ A}(x,n)\wedge{ A}(x,n),\label{2.14}\\
 \a {\cal G}(x,n)= {d} G(x,n) +g_s{G}(x,n)\wedge{G A}(x,n).
\label{2.15}
\ea
In order to obtain the gauge covariant generalized field strength, 
we address the nilpotency of ${\bf d}$
by arranging the following rules: 
\be
         d_{\chi_l}(M_{nk}\chi_k)=(M_{nl}\chi_l)\wedge (M_{nk}\chi_k)=
         M_{nl}M_{lk}\chi_l\wedge\chi_k,
         \label{2.16}
\ee
and in addition,  whenever the $d_{\chi_k}$ operation jumps over 
$M_{nl}\chi_l$, a minus sign is attached.
For example,
\be
    d_{\chi_k}\left(M_{nl}\chi_la(x,n)\right)=
    \left(d_{\chi_k}M_{nl}\chi_l\right)a(x,n)
    - M_{nl}\chi_l\wedge \left(d_{\chi_k}a(x,n)\right).
\label{2.17}
\ee
With these considerations, we can easily calculate: 
\ba
    d_{\chi_l}\left(d_{\chi_k}a(x,n)\right)
  &=&  d_{\chi_l}\left(-a(x,n)M_{nk}\chi_k+M_{nk}\chi_k a(x,n)\right)\nonum
 & =&  -\left(d_{\chi_l}a(x,n)\right)\wedge M_{nk}\chi_k
    -a(x,n)\left(d_{\chi_l}M_{nk}\chi_k\right) 
    \nonum
       && \hskip 0cm   +\left(d_{\chi_l}M_{nk}\chi_k\right)a(x,n)
    -M_{nk}\chi_k \wedge \left( d_{\chi_l}a(x,n)\right)\nonum
 &=&
  -(M_{nl}\chi_l)a(x,n)\wedge(M_{nk}\chi_k)
  +(M_{nl}\chi_l)\wedge(M_{nk}\chi_k)a(x,n)
     \nonum 
  && 
  +(M_{nk}\chi_k)a(x,n)\wedge(M_{nl}\chi_l)
  -(M_{nk}\chi_k)\wedge(M_{nl}\chi_l)a(x,n)
  \label{2.18}
\ea
which yields 
\be
\left(d_{\chi_l}d_{\chi_k}+d_{\chi_k}d_{\chi_l}\right)a(x,n)=0.
\label{2.19}
\ee
In Eq.({\ref{2.18}), we should note that
$\chi_l\wedge\chi_k$ is independent of $\chi_k\wedge\chi_l$ for $k\ne l$
which characterizes the noncommutative property of the NCG in this article. 
It should be emphasized that Eq.(2.19) follows without using the 
shifting rule in Eq.(\ref{2.7}).
From Eq.(\ref{2.19}) we obtain the nilpotency of $d_\chi^{\,2}a(x,n)=0$.
Then,  the nilpotency of ${\bf d}$
\be
{\bf d}^{\,2}a(x,n)=(d^2+dd_\chi+d_\chi d+d_\chi^2)a(x,n)=0     \label{2.20}
\ee
follows with assumptions that $dx^\mu\wedge \chi_k=-\chi_k\wedge dx^\mu$.
Then, we can find Eq.(\ref{2.14}) written as
\be
F(x,n)={\bf d}a_i^\dagger(x,n)\wedge {\bf d}a_i(x,n)+
        A(x,n)\wedge A(x,n) \label{2.14a}
\ee
using the definition of $A(x,n)$ in Eq.(\ref{2.1}).
${\bf d}a_i^\dagger(x,n)\wedge {\bf d}a_i(x,n)$
consists of three parts, of which we investigate 
a two-form with $\chi_k\wedge\chi_l$:
\ba
d_\chi a_i^\dagger(x,n)\wedge d_\chi a_i(x,n)
 &=&\sum_k d_{\chi_k} a_i^\dagger(x,n)\wedge \sum_l d_{\chi_l} a_i(x,n)\nonum
&=&\sum_k\left(-a_i^\dagger(x,n)M_{nk}\chi_k
+M_{nk}\chi_ka_i^\dagger(x,n)\right)\nonum
&&\hskip 2.0cm
\wedge\sum_l\left(-a_i(x,n)M_{nl}\chi_l
+M_{nl}\chi_la_i(x,n)\right).
\label{2.14b}
\ea
Here, we adopt new algebraic rule in order to move $\chi_k$ 
into the right-hand side in the equation. This rule is a extension of
the shifting rule in Eq.(\ref{2.7}) and is applied only in the equation
with more than two-form differentials.
It proceeds as follows:
\ba
d_\chi a_i^\dagger(x,n)\wedge d_\chi a_i(x,n)
&=&\sum_k\sum_{k'}\sum_l\left(-a_i^\dagger(x,n)M_{nk}
                 +M_{nk}a_i^\dagger(x,k')e^{-i\alpha_{kk'}}\right)
\nonum
&&\hskip 1.5cm
 \cdot e^{i\alpha_{kk'}}\left(-a_i(x,k')M_{k'l}
 +M_{k'l}a_i(x,l)\right)\chi_{k'}\wedge\chi_l, 
\label{2.14c}
\ea
where
the sum for $k'$ is taken when $a_i(x,k)=e^{i\alpha_{kk'}}a_i(x,k')$
is satisfied. Thus, $k'$ may be $k$ itself or may be other
different suffix and
$e^{i\alpha_{kk}}=1$. $e^{i\alpha_{kk'}}$ and $e^{-i\alpha_{kk'}}$
emerge when $\chi_k$
jumps over $a_i(x,n)$  and $a_i^\dagger(x,n)$, respectively and 
$\chi_k$ changes to $\chi_{k'}$. 
This means that the gauge field $A_\mu(x,k)$ is equal to $A_\mu(x,k')$ 
or it differs only by the $U(1)$ gauge field. The phase $e^{i\alpha_{kk'}}$
corresponds to that $U(1)$ gauge field. 
This situation is explicitly explained
in the reconstruction of LRSM in the following section.
As a result, we obtain
\ba
d_\chi a_i^\dagger(x,n)\wedge d_\chi a_i(x,n)
&=&\sum_{k,k',l}\left\{ \frac{}{}{\mit\Phi}_{nk}M_{k'l}+M_{nk}{\mit\Phi}_{k'l}
+M_{nk}M_{k'l}\right. \nonum
&&\hskip 2cm \left.\frac{}{}
-a_i^\dagger(x,n)M_{nk}M_{k'l}e^{i\alpha_{kk'}}
a_i(x,l)\right\}\chi_{k'}\wedge\chi_l, \label{2.14d}
\ea
which combined with 
\be
\sum_k{\mit\Phi}_{nk}(x)\chi_k
\wedge\sum_l{\mit\Phi}_{nl}(x)\chi_l
=\sum_{k,k',l}{\mit\Phi}_{nk}(x){\mit\Phi}_{k'l}(x)\chi_{k'}\wedge\chi_l
\label{2.14f}
\ee
in $A(x,n)\wedge A(x,n)$  yields the equation :
\ba
V_{n,k,k',l}(x)&=&\left({\mit\Phi}_{nk}(x)
+M_{nk})({\mit\Phi}_{k'l}(x)+M_{k'l}\right)-
           a_i^\dagger(x,n)M_{nk}M_{k'l}e^{i\alpha_{kk'}}a_i(x,l).
\label{2.14e}
\ea
\par
With these considerations we can find the explicit forms of
the generalized field strength $F(x,n)$ and ${\cal G}(x,n)$.
\ba
  F(x,n) &=& { 1 \over 2}F_{\mu\nu}(x,n)dx^\mu \wedge dx^\nu + 
         \sum_{k\ne n}D_\mu{\mit\Phi}_{nk}(x)dx^\mu \wedge \chi_k \nonum
    && + \sum_{k\ne n}V_{nk}(x)\chi_k \wedge \chi_n 
      + \sum_{k\ne n}\sum_{k'}\sum_{l\ne k'}V_{nkk'l}(x)
                                \chi_{k'} \wedge \chi_l
                \label{2.25}
\ea
where
\ba
 &&  F_{\mu\nu}(x,n)=\partial_\mu A_\nu (x,n) - \partial_\nu A_\mu (x,n) 
               + [A_\mu(x,n), A_\mu(x,n)], \label{2.26}\\
  &&  D_\mu{\mit\Phi}_{nk}(x)=\partial_\mu {\mit\Phi}_{nk}(x) - 
  ({\mit\Phi}_{nk}(x)+M_{nk})A_\mu(x,k)\nonum
    && \hskip 5.5cm + A_\mu(x,n)(M_{nk} + {\mit\Phi}_{nk}(x)),\label{2.27}\\
 && V_{nk}(x)= ({\mit\Phi}_{nk}(x) + M_{nk})({\mit\Phi}_{kn}(x) + M_{kn}) -
             Y_{nk}(x) \hskip 0.5cm{\rm for} \quad k \ne n,
     \label{2.28}\\
 && V_{nkk'l}(x)= ({\mit\Phi}_{nk}(x) + M_{nk})
     ({\mit\Phi}_{k'l}(x) + M_{k'l}) -
      Y_{nkk'l}(x) \hskip 0.5cm{\rm for} \quad k \ne n,\quad l\ne k'.
      \label{2.29}
\ea
$Y_{nk}(x)$ and $Y_{nkk'l}(x)$ are the auxiliary fields written as
\ba
  && Y_{nk}(x)=\sum_i a^\dagger_i(x,n)M_{nk}M_{kn}a_i(x,n), \label{2.30}\\
  && Y_{nkk'l}(x)=\sum_i a^\dagger_i(x,n)
                 M_{nk}M_{k'l}e^{i\alpha_{kk'}}a_i(x,l), \label{2.31}
\ea
which may be a constant or an independent fields.
${\cal G}(x,n)$ is expressed as
\be
   {\cal G}(x,n)=\frac12G_{\mu\nu}(x,n)dx^\mu\wedge dx^\nu, \label{2.32}
\ee
where
\be
       G_{\mu\nu}(x,n)=\p_\mu G_\nu(x,n)-\p_\nu G_\mu(x,n)
       +g_s [G_\nu(x,n), G_\mu(x,n)].     \label{2.33}
\ee
\par
We address the gauge transformation 
of $a_i(x,n)$,  which is defined as 
\ba
&&      a^{g}_{i}(x,n) = a_{i}(x,n)g_f(x,n), \label{2.21} \\
&&      b_j^g(x,n)=b_j(x,n)g_c(x)           \label{2.22}
\ea
where
$g_f(x,n)$ and $g_c(x)$ are the gauge functions 
with respect to the corresponding
flavor unitary group specified by the argument $n$ 
and color group, respectively.
It should be noted that $g_f(x,n)$ and $g_c(x)$ commute with each other. 
Let us define the $d_\chi$
operation on $g_f(x,n)$ by
\ba 
       {\bf d}g_f(x,n)&=&(d+\sum_kd_{\chi_k}) g_f(x,n) \nonum
             & =& \p_\mu g_f(x,n)dx^\mu
             +\sum_k[-g_f(x,n)M_{nk} + M_{nk}g_f(x,k)]\chi_k.
          \label{2.23a}
\ea
Then, we can find from Eq.(\ref{2.1}) 
the gauge transformation of ${\cal A}(x,n)$ to be
\be
{\cal A}^g(x,n)=g^{-1}(x,n) 
{\cal A}(x,n)g(x,n)  +g^{-1}(x,n){\bf d}g(x,n), \label{2.23}
\ee
where $g(x,n)=g_f(x,n)g_c(x)$.
From Eq.(\ref{2.23}), we can find the gauge transformations for 
gauge and Higgs fields as
\ba
&&       A_\mu^g(x,n)=g^{-1}_{_f}(x,n)A_\mu(x,y)g_{_f}(x,n)+
                       g^{-1}_{_f}(x,n)\p_\mu g_{_f}(x,n),  \label{2.23a1}\\
&&       {\mit\Phi}^g_{nm}(x)=g^{-1}_{_f}(x,n){\mit\Phi}_{nm}(x)g_{_f}(x,m)+
                          g^{-1}_{_f}(x,n)\p_{nm}g_{_f}(x,n), \label{2.23b}\\
&&      G_\mu^g(x)=g^{-1}_c(x)G_\mu(x)g_c(x)+
                        \frac1{g_s}g^{-1}_c(x)\p_\mu g_c(x).  \label{2.23c}
\ea
In Eq.(\ref{2.23b}), the operator $\p_{nm}$ is defined as
\be
\p_{nm}g_{_f}(x,n)=-g_{_f}(x,n)M_{nm}+M_{nm}g_{_f}(x,m), \label{2.23ba}
\ee
which seems to be regarded as a difference operator in $Z_{_N}$.
Equation(\ref{2.23b}) is very similar to other two equations and so 
it strongly indicates that the Higgs field is a kind of gauge field
on the discrete space $M_4\times Z_{_N}$. From Eq.(\ref{2.23ba}),
Eq.(\ref{2.23b})  is rewritten as
\be
       {\mit\Phi}^g_{nm}(x)+M_{nm}=g^{-1}_{_f}(x,n)({\mit\Phi}_{nm}(x)+M_{nm})
                  g_{_f}(x,m),
                              \label{2.23bb}\\
\ee
which makes it obvious that 
\be
H_{nm}(x)={\mit\Phi}_{nm}(x)+M_{nm} \label{2.23bc}
\ee
is un-shifted Higgs field whereas ${\mit\Phi}_{nm}(x)$ 
denotes shifted one with the vanishing vacuum expectation value.
From Eqs.(\ref{2.13}) and (\ref{2.23}), 
the gauge-covariant transformation of the generalized field strength
is found to be
\be
         {\cal F}^g(x,n)=g^{-1}(x,n){\cal F}(x,n)g(x,n),  \label{2.24}
\ee
which insures the introduction of the gauge invariant Lagrangian.
\par
In order to construct the Yang-Mills-Higgs Lagrangian.
we define the metric structure of one-forms  as
\be
<dx^\mu, dx^\nu>=g^{\mu\nu},\hskip 1cm 
<\chi_n, dx^\mu>=0, \hskip 1cm
<\chi_n, \chi_k>=-\delta_{nk},
\label{2.34}
\ee
According to these metric structures and Eq.(\ref{2.25}) we can obtain 
the expression for gauge-invariant Yang-Mills-Higgs Lagrangian
\ba
{\cal L}_{{\mathop{}_{\rm YMH}}}(x)
&=&-\sum_{n=1}^{\mathop{}_{N}}{1 \over g_{n}^2}
                  < {\cal F}(x,n),  {\cal F}(x,n)>\nonum
&=&-{\rm Tr}\sum_{n=1}^{\mathop{}_{N}}{1\over 2g^2_n}
                    F_{\mu\nu}^{\dag}(x,n)F^{\mu\nu}(x,n)
             -{\rm Tr}\sum_{n=1}^{\mathop{}_{N}}{1\over 2g^2_n}
               G_{\mu\nu}^{\dag}(x,n)G^{\mu\nu}(x,n)
\nonum
&&+{\rm Tr}\sum_{n=1}^{\mathop{}_{N}}\sum_{k\ne n}{1\over g_{n}^2}
    (D_\mu{\mit\Phi}_{nk}(x))^{\dag}D^\mu{\mit\Phi}_{nk}(x)\nonum
&&\hskip -1cm -{\rm Tr}\sum_{n=1}^{\mathop{}_{N}}
{1\over g_{n}^2}\sum_{k\ne n}V_{nk}^{\dag}(x)V_{nk}(x)
-{\rm Tr}\sum_{n=1}^{\mathop{}_{N}}{1\over g_{n}^2}
    \sum_{k\ne n}\sum_{k'}\sum_{l\ne k'}V_{nkk'l}^{\dag}(x)V_{nkk'l}(x),
\label{2.35}
\ea
where $g_n$ are constants relating to the gauge coupling constant
and
Tr denotes the trace over internal symmetry matrices. 
The last two terms are the potential and interaction terms 
of the Higgs particles. 
\section{Reconstruction of the Yang-Mills-Higgs sector in \lr}
The purpose of this section is to reconstruct the left-right symmetric gauge
model (\lr) based on the formulation in the previous section. 
\lr is the gauge theory with the symmetry 
$SU(3)_c\times SU(2)_{_{}^L}\times SU(2)_{_{}^R}\times  U(1)$ 
and is  promising  as an alternative model of the standard model because
\lr exhibits the seesaw mechanism for the left- and right-handed neutrinos 
\cite{Gell} and
contains two doublet Higgs bosons to give masses to the up and down quarks
separately.
In addition, it is important in the sense that 
the symmetry of \lr is the intermediate symmetry of the 
spontaneously broken SO(10) GUT. 
\par
At first, we must consider the 
formations of gauge fields and the Higgs fields which are incorporated in
the generalized gauge field ${\cal A}(x,n)$. 
Generally speaking, it needs a little bit devices to reproduce the
necessary terms for the Higgs potential and interactions so as to be
explain the phenomenology of \lr, 
whereas the kinetic terms of gauge and Higgs fields come out
without any difficulty. 
In order to meet the consistent presentation of \lr, we consider the 
discrete space $M_4\times Z_4$ on which we specify the generalized 
gauge field $A(x,n)\,(n=1,2,3,4)$ 
and the fermion fields $\psi(x,n)\,(n=1,2,3,4)$. 
Let us begin with the specification of the fermion field  $\psi(x,n)$
on the discrete space $M_4\times Z_4$.
\be
       \psi(x,1)=\frac{1}{\sqrt{2}}\left(\matrix{ 
                                u^r_{_{}^L}\cr
                                u^g_{_{}^L}\cr
                                u^b_{_{}^L}\cr
                                \nu_{_L}\cr
                                d^{\,r}_{_{}^L}\cr
                                d^{\,g}_{_{}^L}\cr
                                d^{\,b}_{_{}^L}\cr
                                e_{_L}\cr }
                            \right), 
       \psi(x,2)=\frac{1}{\sqrt{2}}\left(\matrix{ 
                                u^r_{_{}^R}\cr
                                u^g_{_{}^R}\cr
                                u^b_{_{}^R}\cr
                                 \nu_{_R}       \cr
                                d^{\,r}_{_{}^R}\cr
                                d^{\,g}_{_{}^R}\cr
                                d^{\,b}_{_{}^R}\cr
                                e_{_R}\cr }
                                \right), 
       \psi(x,3)=\frac{1}{\sqrt{2}}\left(\matrix{ 
                                {d^{\,r}_{_{}^L}}^c\cr
                                {d^{\,g}_{_{}^L}}^c\cr
                                {d^{\,b}_{_{}^L}}^c\cr
                                {e_{_L}}^c\cr 
                                -{u^r_{_{}^L}}^c\cr
                                -{u^g_{_{}^L}}^c\cr
                                -{u^b_{_{}^L}}^c\cr
                                -{\nu_{_L}}^c\cr
                                }
                            \right), 
       \psi(x,4)=\frac{1}{\sqrt{2}}\left(\matrix{ 
                                {d^{\,r}_{_{}^R}}^c\cr
                                {d^{\,g}_{_{}^R}}^c\cr
                                {d^{\,b}_{_{}^R}}^c\cr
                                {e_{_R}}^c\cr 
                                -{u^{\,r}_{_{}^R}}^c\cr
                                -{u^{\,g}_{_{}^R}}^c\cr
                                -{u^{\,b}_{_{}^R}}^c\cr
                                -{\nu_{_R}}^c       \cr
                                }
                            \right), \label{3.1}
\ee
where subscripts $L$ and $R$ denote the left-handed and
right-handed fermions, respectively and 
superscripts $r$, $g$ and $b$ represent the color indices. 
The superscript $c$ in $\psi(x,3)$ and $\psi(x,4)$ represents the anti-particle
of the respective fermions.
It should be noticed that $\psi(x,n)$ has the index for the three generation
and so do the explicit expressions for fermions in the right hand sides
of Eq.(\ref{3.1}).  In the strict expressions,
 $u,$ $d,$  $\nu$ and $e$
 in  Eq.(\ref{3.1}) should be replaced by
\be
       u \to \left(\matrix{ u \cr
                            c \cr
                            t\cr}
                            \right), 
                           \hskip 1cm
       d \to \left(\matrix{ d \cr
                            s \cr
                            b \cr}
                            \right), \hskip 1cm
       \nu \to \left(\matrix{ \nu_e  \cr
                              \nu_\mu \cr
                              \nu_\tau\cr}
                            \right), \hskip 1cm
       e \to \left(\matrix{ e  \cr
                            \mu \cr
                            \tau\cr}
                            \right), 
                            \label{3.2}
\ee
respectively. Thus, $\psi(x,n)\,(n=1,2,3,4)$ 
is a vector in the 24-dimensional space.
With these specifications of the fermion fields 
written in 24-dimensional vector, the generalized gauge fields $A(x,n)$ 
must be written in $24\times 24$ matrices to express the interactions 
between fermions and the generalized gauge bosons.
We begin with the construction of the Yang-Mills-Higgs sector of \lr.
\subsection{ Yang-Mills-Higgs Lagrangian}
In order to get Yang-Mills-Higgs Lagrangian in \lr 
we prepare the discrete space $M_4\times Z_4$. 
At first, we specify ${\cal A}(n)\,(n=1,2,3,4)$ 
in Eqs.(\ref{2.8}), (\ref{2.8a}) and (\ref{2.8b}) 
in the explicit forms of gauge and Higgs fields.
${\mit\Phi}_{nk}={\mit\Phi}_{kn}^\dagger$ 
is implied due to $M_{nk}=M_{kn}^\dagger$
as known from Eq.(\ref{2.10}).
For simplicity, the argument $x$ in the Minkowski space $M_4$ 
is abbreviated in the expressions.
\ba 
     {\cal A}(1)=A_{\mu}(1)dx^\mu+G_\mu(1) dx^\mu
      + {\mit\Phi}_{12}\chi_2,
       \label{3.3}
\ea
where
\ba
    &&A_{\mu}(1)=A_{_{}^L\mu}(1)+B_\mu(1)=
  -{i \over 2}\left\{\sum_k\tau_k\otimes 1^4\otimes 1^3 A_{_{}^{L}\mu}^k
   +c\otimes 1^3B_\mu \right\}, \label{3.4}\\
    &&G_\mu(1) = -{i\over 2}\sum_{a=1}^8 1^2\otimes\lambda'^a\otimes1^3
            {G}_\mu^a(x),\label{3.5}\\
     && {\mit\Phi}_{12}=\phi\otimes 1^4\otimes1^3.  \label{3.6}
\ea
In Eq.(\ref{3.4}), $A_{_{}^{L}\mu}^k$, $B_\mu $ are $SU(2)_{_L}$ and 
$U(1)$ gauge fields, respectively, $\tau^k\,(k=1,2,3)$ is the Pauli matrix
and $c$ is the $U(1)$ hypercharge matrix 
corresponding to $\psi(x,1)$ and $\psi(x,2)$ 
in Eq.(\ref{3.1}) expressed in
\be
   c={\rm diag}\left(\frac13,\frac13,\frac13,-1,
    \frac13,\frac13,\frac13,-1\right).    \label{3.7}
\ee
In Eq.(\ref{3.5}),  $G_\mu^a$ is the color gauge field and 
$\lambda'^a$ is $4\times 4$ matrix 
made of the Gell-Mann matrix $\lambda^a$ by adding $0$ components 
to fourth line and column.
\be
   \lambda'^a  =\left(\matrix{  & &  & 0\cr
                                & \lambda^a &  & 0\cr
                                &  &   &0\cr
                              0 & 0 & 0 & 0 \cr}
                      \right).     \label{3.5a}
\ee
In Eq.(\ref{3.6}), $\phi$ is the  field
containing two doublet Higgs bosons with the expression that
\be
\phi=
\left(
\matrix{
         \phi_2^{0} & \phi_1^+ \cr
         \phi_2^-   & \phi_1^0 \cr
}
\right)
. \label{3.8}
\ee
Together with ${\mit\Phi}_{12}$, we must write down  $M_{12}$ 
which determines the symmetry breaking pattern of $SU(2)_{_{}^L}
\times SU(2)_{_{}^R}\times U(1)$.
\be
M_{12}=M_{21}^\dagger =m_{12}\otimes 1^4 \otimes 1^3
=\left(
\matrix{
  \mu_2 & 0 \cr
  0   & \mu_1 \cr
}\right)\otimes 1^4 \otimes 1^3.
\label{3.9}
\ee
$h=\phi+m_{12}$ is transformed as $(2,2^\ast,0)$ under 
$SU(2)_{_{}^L}\times SU(2)_{_{}^R}\times  U(1)$.
\ba 
     {\cal A}(2)=A_{\mu}(2)dx^\mu+G_\mu(2) dx^\mu
      + {\mit\Phi}_{21}\chi_1+{\mit\Phi}_{24}\chi_4,
       \label{3.10}
\ea
where $G_\mu(2)=G_\mu(1)$ and
\ba
    &&A_{\mu}(2)=A_{_{}^L\mu}(2)+B_\mu(2)=
  -{i \over 2}\left\{\sum_k\tau_k\otimes 1^4\otimes 1^3 A_{_{}^{R}\mu}^k
   +c\otimes 1^3B_\mu \right\}, \label{3.11}\\
     && {\mit\Phi}_{21}={\mit\Phi}_{12}^\dagger
           =\phi^\dagger\otimes 1^4\otimes1^3  \label{3.12}\\
     && {\mit\Phi}_{24}= \xi\otimes b
    \otimes1^3, \qquad
    b= \left(\matrix{0& & & \cr
                    &0& & \cr
                    &  &0& \cr
                    & & &1\cr}\right).
      \label{3.13}
\ea
In Eq.(\ref{3.11}), $A_{_{}^{R}\mu}^k\,(k=1.2.3)$ 
are the $SU(2)_{_{}^{R}}$ gauge fields.
$\xi$ is denoted as
\be
\xi = \left(\matrix{ \xi^- & \xi^0 \cr
                             \xi^{--} & -\xi^-\cr}
                             \right).   \label{3.14}
\ee
The matrix ${\rm diag}(0,0,0,1)$ in Eq.(\ref{3.13}) is placed in order to 
avoid the interaction between quarks and $\xi$,
which leads to the hypercharge conservation.
$M_{24}$ prescribes the symmetry breaking of $SU(2)_{_{}^R}\times U(1)$
down to $U(1)_{_{}^Y}$.
\be
M_{24}=m_{24}\otimes b
  \otimes 1^3
=\left(
\matrix{ 0 & M_{_{}^R} \cr
         0 & 0         \cr
}\right)\otimes b
     \otimes 1^3.
     \label{3.15}
\ee
$\xi_{_{}^R}=\xi+m_{24}$ is transformed as $(0, 3, -2)$ under 
$SU(2)_{_{}^L}\times SU(2)_{_{}^R}\times  U(1)$.
\ba 
     {\cal A}(3)=A_{\mu}(3)dx^\mu+G_\mu(3)dx^\mu
      + {\mit\Phi}_{34}\chi_4,
       \label{3.16}
\ea
where $G_\mu(3)$ is given by replacing ${\lambda'^a}$ in Eq.(\ref{3.5})
 by  the complex conjugate $-{\lambda'^a}^\ast$ and so 
\ba
    &&A_{\mu}(3)=A_{_{}^L\mu}(3)+B_\mu(3)=
  -{i \over 2}\left\{\sum_k\tau_k\otimes 1^4\otimes 1^3 A_{_{}^{L}\mu}^k
   -c\otimes 1^3B_\mu \right\}, \label{3.17}\\
    &&G_\mu(3) = {i\over 2}\sum_{a=1}^8 1^2\otimes
    {\lambda'^a}^\ast\otimes1^3{G}_\mu^a(x),\label{3.5b}\\
     && {\mit\Phi}_{34}=
         {\tilde \phi}\otimes 1^4\otimes1^3.  \label{3.18}
\ea
In Eq.(\ref{3.18}), 
\be
{\tilde \phi}=\tau^2\phi^\ast \tau^2=
\left(
\matrix{
         {\phi_1^{0}}^\ast & -\phi_2^+ \cr
         -\phi_1^-   & {\phi_2^0}^\ast \cr
}
\right)
\label{3.19}
\ee
and 
\be
M_{34}=M_{43}^\dagger =\left(
\matrix{
  \mu_1 & 0 \cr
  0   & \mu_2 \cr
}\right)\otimes 1^4 \otimes 1^3=m_{12}\otimes 1^4 \otimes 1^3.
\label{3.20}
\ee
$M_{34}$ is also responsible for the symmetry breaking of 
$SU(2)_{_{}^L}\times SU(2)_{_{}^R}\times U(1)$.
\ba 
     {\cal A}(4)=A_{\mu}(4)dx^\mu+G_\mu(4)dx^\mu
      + {\mit\Phi}_{42}\chi_2+{\mit\Phi}_{43}\chi_3,
       \label{3.21}
\ea
where $G_\mu(4)=G_\mu(3)$ and
\ba
    &&A_{\mu}(4)=A_{_{}^L\mu}(4)+B_\mu(4)=
  -{i \over 2}\left\{\sum_k\tau_k\otimes 1^4\otimes 1^3 A_{_{}^{R}\mu}^k
   -c\otimes 1^3B_\mu \right\}, \label{3.22}\\
     && {\mit\Phi}_{43}={\mit\Phi}_{34}^\dagger
           ={\tilde\phi}^\dagger\otimes 1^4\otimes1^3,  \label{3.23}\\
     && {\mit\Phi}_{42}={\mit\Phi}_{24}^\dagger=
     \xi^\dagger\otimes b
    \otimes1^3.  \label{3.24}
\ea
It should be noted that $-c$ represents the $U(1)$ hypercharge
of $\psi(x,3)$ and $\psi(x,4)$ in Eq.(\ref{3.1}).
\be
M_{42}=M_{24}^\dagger=
\left(\matrix{ 0 & 0  \cr
         M_{_{}^R} & 0 \cr}
     \right)\otimes b
     \otimes 1^3
     \label{3.25}
\ee
We assume $M_{13}=M_{14}=M_{23}=0$
in the above specifications. 
If we would incorporate the adjoint Higgs field $\xi_{_{}^L}$ 
for $SU(2)\ma{L}$ symmetry to 
completely maintain 
the left-right symmetry,
$\xi_{_{}^L}$ would be assigned in ${\mit\Phi}_{13}$.
However, we do not consider the incorporation of $\xi_{_{}^L}$ 
because the symmetry breaking of $SU(2)_{_{}^L}\times U(1)$ down to
$U(1)_{em}$ is caused by the Higgs doublet field $\phi$.
\par
In the above presentations of the explicit form of the generalized
gauge field ${\cal A}(n)$, the following should be remarked that
\be
[A_{_{}^L\mu}(n), B_\mu(n)]=[A_{_{}^L\mu}(n),G_\mu(n)]
=[G_\mu(n), {\mit\Phi}_{nk}]=0, \label{3.26}
\ee
which enables us to consistently reconstruct the \lr Lagrangian.
\par
The gauge transformation functions $g(x,n)\;(n=1,2,3,4)$
in Eq.(\ref{2.23}) are given as
\ba
&&g\,(x,1)=e^{-ia\alpha(x)}g\ma{L}(x)g_c(x),\;e^{-ia\alpha(x)}\in U(1),\;
g\ma{}(x)\in SU(2)_{_{}^L},\;g_c(x)\in SU(3)_c \nonum
&& g\,(x,2)=e^{-ia\alpha(x)}g\ma{R}(x)g_c(x),\;e^{-ia\alpha(x)}\in U(1),\;
g\ma{R}(x)\in SU(2)_{_{}^R},\;g_c(x)\in SU(3)_c \nonum
&&g\,(x,3)=e^{ia\alpha(x)}g\ma{L}(x)g_c^\ast(x),\;e^{ia\alpha(x)}\in U(1),\;
g\ma{L}(x)\in SU(2)_{_{}^L},\;g_c^\ast(x)\in SU(3)_c \nonum
&&g\,(x,4)=e^{ic\alpha(x)}g\ma{R}(x)g_c^\ast(x),\;e^{ia\alpha(x)}\in U(1),\;
g\ma{R}(x)\in SU(2)_{_{}^R},\;g_c^\ast(x)\in SU(3)_c, 
\label{3.27}
\ea
in which it should be remarked that
these gauge functions are taken so as to satisfy 
\be
   [g\ma{L,R}(x), g_c(x)]=[g\ma{L,R}(x), e^{\pm ia\alpha(x)}]
   =[g_c(x),\;e^{\pm ia\alpha(x)}]=0.  \label{3.28}
\ee
\par
With these preparations, we can find the Yang-Mills-Higgs Lagrangian as
follows:
\be
 {\cal L}_{\mathop{}_{ YMH}}(x) = {\cal L}\ma{ GB} 
 + {\cal L}\ma{\rm HK} 
    - {V}_{\mathop{}_{ HP}}, \label{3.29}
\ee
where 
${\cal L}_{\mathop{}_{ GB}}$ and ${\cal L}_{_{ HK}}$ are kinetic terms of
the gauge and Higgs boson fields, respectively and 
${ V}_{_{ HP}}$  contains
the potential and interacting terms of the Higgs bosons.
${\cal L}_{\mathop{}_{GB}}$ is expressed as
\ba
{\cal L}_{\mathop{}_{GB}}&=&
-{1\over 4}\left(
{{12}\over {g_1^2}}+{{12}\over {g_3^2}}\right)
           \sum_{i=1}^3 F^i_{{\mathop{}^{L}}\mu\nu}F_{\mathop{}_{L}}^{i\mu\nu}
-{1\over 4}\left({{12}\over g_2^2}+{{12}\over g_4^3}\right)
\sum_{i=1}^3 F^i_{{\mathop{}^{R}}\mu\nu} F_{\mathop{}^{R}}^{i\mu\nu}\nonum
&&-{1\over4}
\left(\sum_{k=1}^4{3\over {2g_k^2}}{\rm Tr}\:c^2 \right)
B_{\mu\nu}B^{\mu\nu}\nonum
&&-{1\over 4}
  \left(\sum_{k=1}^4{6\over g_k^2}\right) \sum_{a=1}^8
   G_{\mu\nu}^a G^{a\mu\nu}
\label{3.30}
\ea
By defining the gauge coupling constants as
\ba
    &&   \frac 1{g_{\mathop{}_{L}^2}}=\frac{12}{g_1^2}+\frac{12}{g_3^2},
    \label{3.31}\\
    &&   \frac 1{g_{\mathop{}_{R}^2}}=\frac{12}{g_2^2}+\frac{12}{g_4^2},
    \label{3.32}\\
    &&   \frac 1{g_{\mathop{}_{B}^2}}=
               \sum_{k=1}^4{3\over {2g_k^2}}{\rm Tr}\:c^2 ,
    \label{3.33}\\
    &&   \frac 1{g_{s'}^2}=
               \sum_{k=1}^4{6\over {g_k^2}},
    \label{3.34}
\ea
and carrying out the scale transformations of gauge fields, we find           
\ba
{\cal L}_{\mathop{}_{ GB}}&=&
 -{1\over 4}
\sum_i F^i_{{\mathop{}_{L}}\mu\nu} F_{\mathop{}_{L}}^{i\mu\nu}
-{1\over 4}
\sum_i F^i_{{\mathop{}_{R}}\mu\nu} F_{\mathop{}_{R}}^{i\mu\nu}
-{1\over4}B_{\mu\nu}B^{\mu\nu}
-{1\over 4}\sum_a G_{\mu\nu}^a G^{a\mu\nu}
\label{3.35}
\ea
where
\ba
&& F_{{\mathop{}_{L,R}}\mu\nu}^i  = \partial_\mu A_{{\mathop{}_{L,R}}\nu}^i
          -\partial_\nu A_{{\mathop{}_{L,R}}\mu}^i
       +g_{\mathop{}_{L,R}}\,\epsilon^{ijk}A_{{\mathop{}_{L,R}}\mu}^j 
       A_{{\mathop{}_{L,R}
       }\nu}^k,        \label{3.36}\\
&& B_{\mu\nu}=\partial_\mu B_\nu-\partial_\nu B_\mu,\label{3.37}\\
&& G_{\mu\nu}^a  = \partial_\mu G_\nu^a-\partial_\nu G_\mu^a
       +g_cf^{abc}G_\mu^b G_\nu^c. \label{3.38}
\ea
with the definition 
\be
 g_c=g_{s'}g_s=g_{s'}\left(\sum_{k=1}^4{6\over {g_k^2}}\right)^{-1}. 
 \label{3.34a}
\ee
\par
Higgs kinetic term ${\cal L}\ma{HK}$ consists of two terms $L\ma{hK}$ and
$L\ma{\xi\ma{R} K}$ for the Higgs doublet and $SU(2)\ma{R}$ adjoint bosons, 
respectively.  Four kinetic terms of ${\mit\Phi}_{12}$, 
${\mit\Phi}_{21}$, ${\mit\Phi}_{34}$ 
and ${\mit\Phi}_{43}$  contribute to $L\ma{hK}$ and as a result, 
$L\ma{hK}$ is written as 
\be
     L\ma{hK}=12\left({\frac1{g_1^2}}+{\frac1{g_2^2}}
                   +{\frac1{g_3^2}}+{\frac1{g_4^2}}\right)
                   {\rm Tr }\left|D_\mu h\right|^2,  \label{3.39}
\ee
where
\be
      D_\mu h=\p_\mu h-{\frac i2}g\ma{L}\sum_{k=1}^3\tau^k{A\ma{L}^k}_\mu h
                        -{\frac i2}g\ma{R}h\sum_{k=1}^3\tau^k{A\ma{R}^k}_\mu
                        \label{3.40}
\ee
with the definition that
\be
    h=\phi+m_{12}=\left(
 \matrix{
         \phi_2^{0}+\mu_2 & \phi_1^+ \cr
         \phi_2^-   & \phi_1^0+\mu_1 \cr
}
\right)
. \label{3.41}
\ee
The rescaling of $\sqrt{\gamma_h }h \to h$ with 
\be
\gamma_h=12\left({\frac1{g_1^2}}+{\frac1{g_2^2}}
                   +{\frac1{g_3^2}}+{\frac1{g_4^2}}\right) \label{3.42}
\ee
leads  Eq.(\ref{3.39}) to the standard form of the kinetic term :
\be
     L\ma{hK}={\rm Tr }\left|D_\mu h\right|^2  \label{3.43}
\ee
with the same expression for $D_\mu h$ as in Eq.(\ref{3.40}).
The kinetic term of the $SU(2)\ma{R}$ adjoint Higgs boson $\xi\ma{R}$
is given as 
\be
    L\ma{\xi\ma{R} K} = 3\left({\frac1{g_2^2}}+{\frac1{g_4^2}}\right)
             {\rm Tr}\left|D_\mu \xi\ma{R}\right|^2, \label{3.44}
\ee
where
\be
   D_\mu\xi\ma{R}=\p_\mu \xi\ma{R}-\frac i2g\ma{R}\sum_{k=1}^3
         \left[\tau^kA\ma{R}^k, \xi\ma{R}\right] + ig\ma{B}B_\mu\xi\ma{R}
                \label{3.45}
\ee
with the notation that
\be
           \xi\ma{R}=\xi+m_{24}=
           \left(
 \matrix{
         \xi^{-} & \xi^{0}+M\ma{R} \cr
         \xi^{--}   & -\xi^{-} \cr}
          \right). \label{3.46}
\ee
The rescaling of $\sqrt{\gamma_\xi} \xi\ma{R} \to \xi\ma{R}$ with 
\be
\gamma_\xi=3\left({\frac1{g_2^2}}+{\frac1{g_4^2}}\right) \label{3.47}
\ee
leads Eq.(\ref{3.44}) to the standard form :
\be
    L\ma{\xi\ma{R} K} = {\rm Tr}\left|D_\mu \xi\ma{R}\right|^2
                 \label{3.48}
\ee
with the same expression for $D_\mu \xi\ma{R}$ as in Eq.(\ref{3.45}).
\par
Then, we move to the Higgs potential and interacting terms which are 
explained somewhat in detail because the managements of the auxiliary 
fields in Eqs.(\ref{2.30}) and (\ref{2.31}) are rather complicated.
For the simplicity, $g_1=g_3$ and $g_2=g_4$ are assumed, which is plausible
from the explicit assignments of fields in Eqs.(\ref{3.3}), (\ref{3.10}),
(\ref{3.16}) and (\ref{3.21}).
$V\ma{HP}$ in Eq.(\ref{3.29}) consists of six terms as
\be
V\ma{HP}=V_{hh^\dagger}+V_{h{\tilde h}}+V_{h\xi\ma{R}}
+V_{\xi\ma{R}\xi\ma{R}^\dagger}+V'_{\xi\ma{R}\xi\ma{R}^\dagger}
+V_{\xi\ma{R}\xi\ma{R}}, \label{3.51}
\ee
which are explicitly expressed in order. 
After the rescaling of $\gamma_hh\to h$, $V_{hh^\dagger}$ is written as
\ba
V_{hh^\dagger}&=&
         \frac{12}{g_1^2\gamma_h^2}
         {\rm Tr}\left|hh^\dagger-y_{12}\right|^2
        + \frac{12}{g_2^2\gamma_h^2}
        {\rm Tr}\left|h^\dagger h-y_{21}\right|^2\nonum
  &&\hskip 2cm   + \frac{12}{g_1^2\gamma_h^2}{\rm Tr}
         \left|{\tilde h}{\tilde h}^\dagger-y_{34}\right|^2
        + \frac{12}{g_2^2\gamma_h^2}{\rm Tr}
         \left|{\tilde h}^\dagger{\tilde h}-y_{43}\right|^2, \label{3.52}
\ea
where the auxiliary fields are written in $2\times2$ matrices by taking 
the trace of the common factor $1^4\otimes1^3$ and expressed as
\ba
  &&  y_{12}=\sum_i a_i^\dagger(1)m_{12}m_{21}a_i(1)
                =\sum_i a_i^\dagger(1) 
                  \left(\matrix{\mu_2^2  & 0       \cr
                                0       & \mu_1^2 \cr}\right)
                                 a_i(1),\label{3.53}\\
  &&  y_{21}=\sum_i a_i^\dagger(2)m_{21}m_{12}a_i(2)
               =\sum_i a_i^\dagger(2) 
                  \left(\matrix{\mu_2^2  & 0       \cr
                                0       & \mu_1^2 \cr}\right)
                                a_i(2),\label{3.54}\\
  &&  y_{34}=\sum_i a_i^\dagger(3)m_{34}m_{43}a_i(3)
            =\sum_i a_i^\dagger(3) 
                  \left(\matrix{\mu_1^2  & 0      \cr
                                0       & \mu_2^2 \cr}\right)
                                a_i(3),\label{3.55}\\
  &&  y_{43}=\sum_i a_i^\dagger(4)m_{43}m_{34}a_i(4)
                =\sum_i a_i^\dagger(4) 
                  \left(\matrix{\mu_1^2  & 0      \cr
                                0       & \mu_2^2 \cr}\right)
                                a_i(4).\label{3.56}
\ea
In these equations, $a_i(n)\;(n=1,2,3,4)$ are reduced to $2\times2$ matrices
after taking the trace of $1^4\otimes1^3$.  
In managing these auxiliary fields, we should remark that 
from the explicit assignments of gauge fields 
in Eqs.(\ref{3.4}), (\ref{3.11}),
(\ref{3.17}) and (\ref{3.22}), only difference of 
$A_\mu(1)\:(A_\mu(2))$ and 
$A_\mu(3)\:(A_\mu(4))$ is only the contribution of the $U(1)$ gauge field.
Therefore, we can conjecture from the constituent form of the gauge field in
Eq.(\ref{2.9}) that the relations
\be
    a_i(1)=e^{i\alpha_i}a_i(3),\hskip 2cm a_i(2)=e^{i\alpha_i}a_i(4) 
                   \label{3.57}    
\ee
hold for fundamental fields $a_i(n)\,(n=1,2,3,4)$.
Owing to Eqs.(\ref{2.1a}) and (\ref{3.57}), we can find the relations
\be
  y_{12}+y_{34}=(\mu_1^2+\mu_2^2)\cdot1^2, \hskip 2cm 
  y_{21}+y_{43}=(\mu_1^2+\mu_2^2)\cdot1^2.
  \label{3.58}
\ee
Eliminating the independent auxiliary field $y_{12}$ in Eq.(\ref{3.52})
by use of the equation of motion,  we obtain
\ba
   V_{hh^\dagger}&=&
    \left({\frac6{g_1^2\gamma_h^2}}+{\frac6{g_2^2\gamma_h^2}}\right)
   {\rm Tr} \left|hh^\dagger+{\tilde h}{\tilde h}^\dagger
                    -(\mu_1^2+\mu_2^2)\cdot1^2\right|^2\nonum
          &&+\frac{24}{g_2^2\gamma_h^2}
   {\rm Tr}\left|y_{212}-\frac{h^\dagger h-{\tilde h}^\dagger{\tilde h}
          +(\mu_1^2+\mu_2^2)\cdot 1^2}{2}\right|^2,
                     \label{3.59}
\ea
where the last term is later combined 
with the term occurring in $V_{\xi\ma{R}\xi\ma{R}^\dagger}$.
\par
With the same rescaling of $\gamma_hh\to h$, $V_{h{\tilde h}}$
is written as
\ba
 V_{h{\tilde h}}&=&
   \frac{12}{g_1^2\gamma_h^2}
   {\rm Tr}\left|h{\tilde h}^\dagger-y_{1243}\right|^2
  +\frac{12}{g_2^2\gamma_h^2}
  {\rm Tr}\left|h^\dagger{\tilde h}-y_{2134}\right|^2\nonum
 &&\hskip 2cm 
    +\frac{12}{g_1^2\gamma_h^2}
    {\rm Tr}\left|{\tilde h}h^\dagger-y_{3421}\right|^2
  +\frac{12}{g_2^2\gamma_h^2}
  {\rm Tr}\left|{\tilde h}^\dagger h-y_{4312}\right|^2,
\label{3.60}
\ea
where        
\ba
  &&  y_{1243}=\sum_i a_i^\dagger(1)m_{12}m_{43}e^{i\alpha_i}a_i(3)
               =\sum_i a_i^\dagger(1) 
                  \left(\matrix{\mu_1\mu_2  & 0       \cr
                                0       & \mu_1\mu_2 \cr}\right)
                                 e^{i\alpha_i}a_i(3),\label{3.61}\\
  &&  y_{2134}=\sum_i a_i^\dagger(2)m_{21}m_{34}e^{i\alpha_i}a_i(4)
             =\sum_i a_i^\dagger(2) 
                  \left(\matrix{\mu_1\mu_2  & 0       \cr
                                0       & \mu_1\mu_2 \cr}\right)
                               e^{i\alpha_i} a_i(4),\label{3.62}\\
  &&  y_{3421}=\sum_i a_i^\dagger(3)m_{34}m_{21}e^{-i\alpha_i}a_i(1)
               =\sum_i a_i^\dagger(3) 
                  \left(\matrix{\mu_1\mu_2  & 0      \cr
                                0       & \mu_1\mu_2 \cr}\right)
                               e^{-i\alpha_i} a_i(1),\label{3.63}\\
  &&  y_{4312}=\sum_i a_i^\dagger(4)m_{43}m_{12}e^{-i\alpha_i}a_i(2)
              =\sum_i a_i^\dagger(4) 
                  \left(\matrix{\mu_1\mu_2  & 0      \cr
                                0       & \mu_1\mu_2 \cr}\right)
                                e^{-i\alpha_i}a_i(2).\label{3.64}
\ea
According to Eq.(\ref{3.57}) together with Eq.(\ref{2.1a}),
we find
\be
y_{1243}=y_{2134}=y_{3421}=y_{4312}=\mu_1\mu_2\cdot 1^2. \label{3.65}
\ee
Inserting Eq.(\ref{3.65}) into Eq.(\ref{3.60}), we obtain
\be
     V_{h{\tilde h}}=
     \left({\frac2{g_1^2\gamma_h^2}}+{\frac2{g_2^2\gamma_h^2}}\right)
{\rm Tr}\left|h{\tilde h}^\dagger-\mu_1\mu_2\cdot1^2\right|^2 \label{3.66}
\ee
because of ${\rm Tr}|h{\tilde h}^\dagger-\mu_1\mu_2\cdot1^2|^2
={\rm Tr}|h^\dagger{\tilde h}-\mu_1\mu_2\cdot1^2|^2$.
\par
Then, let us move to $V_{h\xi\ma{R}}$. After the rescaling of 
$\gamma_hh\to h$ and $\gamma_\xi\xi\ma{R} \to \xi\ma{R}$, 
$V_{h\xi\ma{R}}$ is written as
\ba
 V_{h\xi\ma{R}}&=&
      \frac3{g_1^2\gamma_h\gamma_\xi}\left\{\left| 
 h\xi\ma{R}-y_{1224}\right|^2+\left|h\xi\ma{R}^\dagger-y_{1242}\right|^2
      \right\}\nonum
      &&+
      \frac3{g_2^2\gamma_h\gamma_\xi}\left\{\left| 
      \xi\ma{R}{\tilde h}^\dagger-y_{2443}\right|^2
      +\left|\xi\ma{R}h^\dagger-y_{2421}\right|^2
      \right\}\nonum
      &&+
      \frac3{g_3^2\gamma_h\gamma_\xi}\left\{\left| 
      {\tilde h}\xi\ma{R}^\dagger-y_{3442}\right|^2
      +\left|{\tilde h}\xi\ma{R}-y_{3424}\right|^2
      \right\}\nonum
      &&+
      \frac3{g_4^2\gamma_h\gamma_\xi}\left\{\left| 
      \xi\ma{R}^\dagger h^\dagger-y_{4221}\right|^2
      +\left|\xi\ma{R}^\dagger{\tilde h}^\dagger-y_{4243}\right|^2
      \right\}, \label{3.67}
\ea      
where
\ba
&&  y_{1224}=\sum_ia^\dagger(1)m_{12}m_{24}a_i(4)
         =\sum_ia^\dagger(1)
          \left(\matrix{0 & \mu_2M\ma{R}\cr
                        0 &  0          \cr}\right)
          a_i(4),   \label{3.68}\\
&&  y_{1242}=\sum_ia^\dagger(1)m_{12}m_{42}a_i(2)
         =\sum_ia^\dagger(1)
          \left(\matrix{0            & 0\cr
                        \mu_1M\ma{R} &  0  \cr}\right)
          e^{i\alpha_i}a_i(2),   \label{3.69}\\
&&  y_{2443}=\sum_ia^\dagger(2)m_{24}m_{43}a_i(3)
         =\sum_ia^\dagger(1)
          \left(\matrix{0 & \mu_2M\ma{R}\cr
                        0 &  0          \cr}\right)
          a_i(4),   \label{3.70}\\
&&  y_{2421}=\sum_ia^\dagger(2)m_{24}m_{21}a_i(1)
         =\sum_ia^\dagger(2)
          \left(\matrix{0 & \mu_1M\ma{R}\cr
                        0 &  0          \cr}\right)
         e^{-i\alpha_i} a_i(1),   \label{3.71}\\
&&  y_{3442}=\sum_ia^\dagger(3)m_{34}m_{42}a_i(2)
         =\sum_ia^\dagger(1)
          \left(\matrix{0            & 0\cr
                        \mu_2M\ma{R} &  0  \cr}\right)
          a_i(2),   \label{3.72}\\
&&  y_{3424}=\sum_ia^\dagger(3)m_{34}m_{24}a_i(4)
         =\sum_ia^\dagger(3)
          \left(\matrix{0 & \mu_1M\ma{R}\cr
                        0 &  0          \cr}\right)
         e^{-i\alpha_i} a_i(4),   \label{3.73}\\
&&  y_{4221}=\sum_ia^\dagger(4)m_{42}m_{21}a_i(1)
         =\sum_ia^\dagger(4)
          \left(\matrix{0            & 0\cr
                        \mu_2M\ma{R} &  0  \cr}\right)
          a_i(1),   \label{3.74}\\
&&  y_{4243}=\sum_ia^\dagger(4)m_{42}m_{43}a_i(3)
         =\sum_ia^\dagger(4)
          \left(\matrix{0            & 0\cr
                        \mu_1M\ma{R} &  0  \cr}\right)
          e^{i\alpha_i}a_i(1).   \label{3.75}
\ea
According to Eq.(\ref{3.58}) and by rewriting the above equations 
in terms of only $a_i(1)$ and $a_i(2)$, $V_{h\xi\ma{R}}$ is
expressed as
\ba
V_{h\xi\ma{R}}&=& \left({\frac1{g_1^2}}+{\frac1{g_2^2}}\right)
               {\frac3{\gamma_h\gamma_\xi}}
            \left\{\left|h\xi\ma{R}-y_{1224}\right|^2
             +\left| {\tilde h} \xi\ma{R}-y_{3424}\right|^2\right.\nonum
      &&  \hskip 3cm\left. +\left|{h}\xi\ma{R}^\dagger-y_{1242}\right|^2
             +\left|{\tilde h}\xi\ma{R}^\dagger -y_{3442}\right|^2
             \right \},   \label{3.76}
\ea
where
\ba
&& \hskip -0.5cm y_{1224}  =\sum_ia^\dagger(1)
          \left(\matrix{0 & \mu_2M\ma{R}\cr
                        0 &  0          \cr}\right)
          e^{-i\alpha_i}a_i(2),  \quad
  y_{3424} =\sum_ia^\dagger(1)
          \left(\matrix{0 & \mu_1M\ma{R}\cr
                        0 &  0          \cr}\right)
         e^{-i\alpha_i} a_i(2),   \label{3.77}\\
&& \hskip -0.5cm y_{1242}  =\sum_ia^\dagger(1)
          \left(\matrix{0            & 0\cr
                        \mu_1M\ma{R} &  0  \cr}\right)
          e^{i\alpha_i}a_i(2),   \quad
    y_{3442} =\sum_ia^\dagger(1)
          \left(\matrix{0            & 0\cr
                        \mu_2M\ma{R} &  0  \cr}\right)
          e^{i\alpha_i}a_i(2),   \label{3.78}
\ea
From Eqs.(\ref{3.77}) and (\ref{3.78}), we find the relations
$\mu_1y_{1224}=\mu_2y_{3424}$ and $\mu_1y_{3442}=\mu_2y_{1242}$ 
which yield with the aid of equation of motion
\be
V_{h\xi\ma{R}}= \left({\frac1{g_1^2}}+{\frac1{g_2^2}}\right)
               {\frac3{\gamma_h\gamma_\xi}}
 \left[ \frac{{\rm Tr}\left|(\mu_1h-\mu_2{\tilde h})\xi\ma{R}\right|^2
    +{\rm Tr}\left|(\mu_2h-\mu_1{\tilde h})\xi\ma{R}^\dagger\right|^2}
    {\mu_1^2+\mu_2^2}\right]. \label{3.79}
\ee
\par
$V_{\xi\ma{R}\xi\ma{R}^\dagger}$ is written as
\be
V_{\xi\ma{R}\xi\ma{R}^\dagger}=\frac{3}{g_2^2\gamma_\xi^2}
  \left\{ {\rm Tr}\left|\xi\ma{R}\xi\ma{R}^\dagger - y_{24} \right|^2
      + {\rm Tr}\left|\xi\ma{R}^\dagger\xi\ma{R} - y_{42} \right|^2
      \right\}, \label{3.80}
\ee
where
\ba
  && y_{24}=\sum_ia_i^\dagger(2)m_{24}m_{42}a_i(2)=
  \sum_ia_i^\dagger(2)
    \left( \matrix{ M\ma{R}^2 & 0 \cr
                       0      & 0 \cr}\right)
  a_i(2), \label{2.81}\\
  && y_{42}=\sum_ia_i^\dagger(4)m_{42}m_{24}a_i(4)=
  \sum_ia_i^\dagger(4)
    \left( \matrix{    0      & 0 \cr
                       0      & M\ma{R}^2 \cr}\right)
  a_i(4). \label{3.82}  
\ea  
From Eqs.(\ref{3.57}) and (\ref{2.1a}), 
we readily know $y_{24}+y_{42}=M\ma{R}^2\cdot1^2$ 
which yields the equation
\ba
V_{\xi\ma{R}\xi\ma{R}^\dagger}&=&\frac{3}{2g_2^2\gamma_\xi^2}
    {\rm Tr}\left|\xi\ma{R}\xi\ma{R}^\dagger+\xi\ma{R}^\dagger\xi\ma{R}
                   -M\ma{R}^2\cdot1^2 \right|^2 \nonum
  && +\frac{6}{g_2^2\gamma_\xi^2}
    {\rm Tr}\left| y_{242}+
    \frac{\xi\ma{R}\xi\ma{R}^\dagger-\xi\ma{R}^\dagger\xi\ma{R}
                   -M\ma{R}^2\cdot1^2}{2} \right|^2, \label{3.83}
\ea
where the last term is combined with the last term in Eq.(\ref{3.59}).
From Eq.(\ref{2.1a}), we know the relation
\be
   \frac{y_{21}-\mu_2^2}{\mu_1^2-\mu_2^2}+\frac{y_{24}}{M\ma{R}^2}=1^2
 \label{3.84}
\ee
which together with the equation of motion helps us introduce the equation 
\ba
V'_{h\xi\ma{R}}&=&\frac{6}{g_2^2\gamma_\xi^2}
 \frac{1}{1+\displaystyle{\frac{4\gamma_\xi^2}{\gamma_h^2}
           \left(\frac{\mu_2^2-\mu_1^2}{M\ma{R}^2}\right)^2}}
           {\rm Tr}\left| h^\dagger h-{\tilde h}^\dagger{\tilde h}-
           (\mu_2^2-\mu_1^2)
           \left(\matrix{ 1 & 0 \cr
                         0 & -1\cr}\right)\right. \nonum
       && \hskip 4cm \left. -
       \frac{\mu_2^2-\mu_1^2}{M\ma{R}^2}\left\{
       \xi\ma{R}\xi\ma{R}^\dagger-\xi\ma{R}^\dagger\xi\ma{R}-M\ma{R}^2
        \left(\matrix{ 1 & 0 \cr
                         0 & -1\cr}\right)\right\}\right|^2.
                         \label{3.85}
\ea
In the limit of $M\ma{R} >\!> \mu_1,\; \mu_2$, this term is reduced to
\ba
V'_{h\xi\ma{R}}&=&\frac{6}{g_2^2\gamma_\xi^2}
           {\rm Tr}\left| h^\dagger h-{\tilde h}^\dagger{\tilde h}-
           (\mu_2^2-\mu_1^2)
           \left(\matrix{ 1 & 0 \cr
                         0 & -1\cr}\right)\right|^2.
                         \label{3.86}
\ea
It should be remarked that the invariance of Eq.(\ref{3.85}) 
for $SU(2)\ma{R}$ symmetry is
manifest, whereas  Eq.(\ref{3.86}) is not gauge invariant for $SU(2)\ma{R}$.
However, Eq.(\ref{3.86}) is invariant for $SU(2)\ma{L}$ gauge
transformation. Thus, we adopt this approximation hereafter.
As a result of the elimination 
of  auxiliary fields $y_{21}$ and $y_{24}$,
we find the potential terms for $h$ and $\xi\ma{R}$ written as
\ba
 &&  V_{hh^\dagger}=
    \left({\frac6{g_1^2\gamma_h^2}}+{\frac6{g_2^2\gamma_h^2}}\right)
   {\rm Tr} \left|hh^\dagger+{\tilde h}{\tilde h}^\dagger
                    -(\mu_1^2+\mu_2^2)\cdot1^2\right|^2
                     \label{3.87}\\
&& V_{\xi\ma{R}\xi\ma{R}^\dagger}=\frac{3}{2g_2^2\gamma_\xi^2}
    {\rm Tr}\left|\xi\ma{R}\xi\ma{R}^\dagger+\xi\ma{R}^\dagger\xi\ma{R}
                   -M\ma{R}\cdot1^2 \right|^2. 
                   \label{3.88}
\ea
\par
Then, the term $V_{\xi\ma{R}\xi\ma{R}}$ is written as
\be
   V_{\xi\ma{R}\xi\ma{R}}=\frac{3}{g_2^2\gamma_\xi^2}
      \left\{ {\rm Tr}\left|\xi\ma{R}\xi\ma{R}-y_{2424}\right|^2+
      {\rm Tr}\left|\xi\ma{R}^\dagger\xi\ma{R}^\dagger
      -y_{4242}\right|^2 \right\},
  \label{3.89}
\ee
where
\be
    y_{2424}=\sum_ia_i^\dagger(2)m_{24}m_{24}e^{-i\alpha_i}a_i(4)=0,\quad
    y_{4242}=\sum_ia_i^\dagger(4)m_{42}m_{42}e^{i\alpha_i}a_i(2)=0.
\label{3.90}
\ee
These equations simplify Eq.(\ref{3.88}) to yield
\be
  V_{\xi\ma{R}\xi\ma{R}}=\frac{6}{g_2^2\gamma_\xi^2}
     {\rm Tr} \left|\xi\ma{R}\xi\ma{R}\right|^2 \label{3.91}
\ee
which is very important for the doubly charged Higgs boson $\xi^{--}$
to be massive, as shown later.
\par
\subsection{Gauge boson mass}
Let us investigate the various aspects of our LRSM introduced from now.
Gauge boson mass matrix can be extracted from  kinetic terms of 
the Higgs bosons $h$ and $\xi\ma{R}$. 
According to Eqs.(\ref{3.43}) and (\ref{3.48}), 
the following mass matrix for the neutral gauge boson follows. 
\be
M\ma{NGB}=\left(\matrix{
       g\ma{L}^2\mu^2        &   -g\ma{L}g\ma{R}\mu^2  &   0   \cr
       -g\ma{L}g\ma{R}\mu^2   & g\ma{R}^2(\mu^2+M\ma{R}^2) &
      -g\ma{R}g\ma{B}M\ma{R}^2   \cr
      0   &   -g\ma{R}g\ma{B}M\ma{R}^2  & g\ma{B}^2M\ma{R}^2\cr}\right),
      \label{3.92}
\ee
where $\mu^2=(\mu_1^2+\mu_2^2)/2$ 
and the rescaling of $\sqrt{2}M\ma{R}\to M\ma{R}$ is carried out.
Eigenvalues of this mass matrix are easily calculated with the
approximation of $M\ma{R}>\!> \mu$ to be
\ba
&&         M_\gamma^2=0,                   \label{3.93}\\
&&         M\ma{Z}^2=
  \left( g\ma{L}^2 +
  {g\ma{R}^2g\ma{B}^2\over g\ma{R}^2+g\ma{B}^2}\right)
     \mu^2,                                 \label{3.94}\\
&&         M\ma{Z\ma{R}}^2=(g\ma{R}^2+g\ma{B}^2)M\ma{R}^2, 
                                            \label{3.95}
\ea
where
$\gamma$ and  $Z$ represent of course photon and neutral
 weak boson, respectively
and ${Z\ma{R}}$ denotes the extra neutral gauge boson 
expected in this model. 
As a result of the spontaneous breakdown of gauge symmetry in the flavor
sector, only remaining symmetry is the $U(1)$ electromagnetic one.
This is reflected on the vanishing mass of photon field in Eq.(\ref{3.93}).
If we define the coupling constant $g'$ as
\be
     g'^2=\frac{g\ma{R}^2g\ma{B}^2}{g\ma{R}^2+g\ma{B}^2},\label{3.96}
\ee
the mass of the neutral gauge boson $Z_\mu$ is written as
\be
      M\ma{Z}^2=( g\ma{L}^2 + g'^2)\mu^2,        \label{3.97}
\ee
which is a favorite form in accord with the standard model.
In this sense, $g'$ is a coupling constant of the gauge
symmetry resulting from the spontaneous breakdown 
of $SU(2)\ma{R}\times U(1)$ down to $U(1)\ma{Y}$.
$M_{Z\ma{R}}$ is estimated to be so
large that one can not detect $Z\ma{R}$ in the energy range of
the accelerator available nowadays.
\par
The unitary matrix $U$ to transform $({A\ma{L}^3}_\mu, 
{A\ma{R}^3}_\mu, B_\mu)$
into $(A_\mu, Z_\mu, {Z\ma{R}}_\mu)$ is given as
\be
  U=\left(\matrix{  
      \sin\theta\ma{W} &\cos\theta\ma{W}\sin\theta\ma{Y}   
                      & \cos\theta\ma{W}\cos\theta\ma{Y}  \cr
 \cos\theta\ma{W} & -\sin\theta\ma{W}\sin\theta\ma{Y} 
                                  & -\sin\theta\ma{W}\cos\theta\ma{Y}\cr
       0 & \cos\theta\ma{Y}            & -\sin\theta\ma{Y}\cr}
                \right), \label{3.98}
\ee
where
\ba
  &&   \sin\theta\ma{W}=\frac{g'}{\sqrt{g\ma{L}^2+g'^2}},\quad
       \cos\theta\ma{W}=\frac{g\ma{L}}{\sqrt{g\ma{L}^2+g'^2}}, 
       \label{3.99}\\
  &&   \sin\theta\ma{Y}=\frac{g\ma{B}}{\sqrt{g\ma{R}^2+g\ma{B}^2}},\quad
  \cos\theta\ma{Y}=\frac{g\ma{R}}{\sqrt{g\ma{R}^2+g\ma{B}^2}}. 
  \label{3.100}
\ea
$\theta\ma{W}$ is the Weinberg angle and $\theta\ma{Y}$ is
a mixing angle between the gauge bosons ${A\ma{R}^3}_\mu$ and $B_\mu$.
From Eq.(\ref{3.98}), we can denote the transformation
 relations for neutral gauge bosons as
\ba
&& A_\mu={A\ma{L}^3}_\mu\sin\theta\ma{W}
    +{A\ma{R}^3}_\mu\cos\theta\ma{W}\sin\theta\ma{Y}
    + B_\mu\cos\theta\ma{W}\cos\theta\ma{Y}, \label{3.101}\\
&& Z_\mu={A\ma{L}^3}_\mu\cos\theta\ma{W}
    -{A\ma{R}^3}_\mu\sin\theta\ma{W}\sin\theta\ma{Y}
     -B_\mu\sin\theta\ma{W}\cos\theta\ma{Y}, \label{3.102}\\
&& {Z\ma{R}}_\mu={A\ma{R}^3}_\mu\cos\theta\ma{Y}-B_\mu\sin\theta\ma{Y}.
      \label{3.103}
\ea
If $B'_\mu$ is defined as 
$B'_\mu=B_\mu\cos\theta\ma{Y}+{A\ma{R}^3}_\mu\sin\theta\ma{Y}$
we find the well-known mixing relations written as
\ba
&&  A_\mu=B'_\mu\cos\theta\ma{W}
    +{A\ma{L}^3}_\mu\sin\theta\ma{W},\label{3.104}\\
&&  Z_\mu=-B'_\mu\sin\theta\ma{W}
   +{A\ma{L}^3}_\mu\cos\theta\ma{W}\label{3.105}
\ea
and 
\ba
&&  B'_\mu=B_\mu\cos\theta\ma{Y}
     +{A\ma{R}^3}_\mu\sin\theta\ma{Y},\label{3.106}\\
&&  {Z\ma{R}}_\mu=-B_\mu\sin\theta\ma{Y}
   +{A\ma{R}^3}_\mu\cos\theta\ma{Y}.\label{3.107}
\ea
Also from these equations, 
$B'_\mu$ is known to be a gauge boson due to the spontaneous
breakdown of $SU(2)\ma{R}\times U(1)$ down to $U(1)\ma{Y}$.
In this context, we investigate the $U(1)\ma{Y}$ hypercharge assignment 
of two-Higgs doublet bosons in $h$ expressed in Eq.(\ref{3.41}).
From Eq.(\ref{3.40}), we find
\ba
D_\mu h&=&\p_\mu h-\frac{i}2g\ma{L}\sum_{k=1}^3\tau^kA\ma{L}^k\nonum
&& -\frac{i}2 h \left(\matrix{ -g\ma{R}{A\ma{R}^3}_\mu & 0 \cr
                                  0    & g\ma{R}{A\ma{R}^3}_\mu \cr}
                                  \right)
   +\frac{i}2 h \left(\matrix{ 0 & g\ma{R}\sqrt{2}{W\ma{R}^+}_\mu \cr
                           g\ma{R}\sqrt{2}{W\ma{R}^-}_\mu    & 0 \cr}
                                  \right).\label{3.108}
\ea
\par
If we denote $h=(h_2,h_1)$, the $U(1)\ma{Y}$ 
hypercharges of $h_1$ and $h_2$ are
$+1$ and $-1$ from Eq.(\ref{3.108}), respectively because 
\ba
g\ma{R}{A\ma{R}^3}_\mu
&=&g\ma{R}(B'_\mu\sin\theta\ma{Y}+{Z\ma{R}}_\mu\cos\theta\ma{Y})
=g'B'_\mu+g\ma{R}{Z\ma{R}}_\mu\cos\theta\ma{Y} \label{3.109}
\ea
due to Eq.(\ref{3.100}).
\par
Then,  the mass matrix for charged gauge bosons 
is given also from Eqs.(\ref{3.43}) and (\ref{3.48}) as
\be
M\ma{CGB}=\left(\matrix{
       g\ma{L}^2\mu^2 &   -g\ma{L}g\ma{R}\mu_1\mu_2          \cr
       -g\ma{L}g\ma{R}\mu_1\mu_2       &   
       g\ma{R}^2(\mu^2+\displaystyle{\frac12}M\ma{R}^2) \cr
      }\right). \label{3.110}
\ee
In the limit of $M\ma{R} >\!> \mu$,  the eigenvalues are written as
\ba
   &&  M\ma{W}^2=g\ma{L}^2\mu^2, \label{3.111}\\
   &&  M\ma{W\ma{R}}^2=\frac12g\ma{R}^2M\ma{R}^2,
\label{3.112}
\ea
which together with the investigation of eigenstates indicates
that the mixing 
between ${W\ma{L}^{\pm}}_\mu$ and ${W\ma{R}^{\pm}}_\mu$
does not take place.
$W_\mu^\pm={W\ma{L}^{\pm}}_\mu$ is the charged weak boson and
$W\ma{R}^\pm={W\ma{R}^{\pm}}_\mu$ 
is also the extra charged gauge boson expected in this 
model and its mass is so high that it can not be detectable.
Combined Eqs.(\ref{3.111}) with (\ref{3.97}), the mass relation
\be
  M\ma{W}=M\ma{Z}\cos\theta\ma{W} \label{3.113}
\ee
holds. This is the same relation as in the standard model.
\par
Then, we address the numerical values of the coupling constants 
and mixing angles $\theta\ma{W}$ and $\theta\ma{Y}$ within 
tree level when $g_1=g_2=g_3=g_4$.
From Eqs.(\ref{3.31})$\sim$(\ref{3.34}) 
and with the remark of ${\rm Tr}\: c^2=8/3$, the following equations hold:
\be
g\ma{L}^2=\frac{g_1^2}{24}, \quad g\ma{R}^2=\frac{g_1^2}{24},\quad
g\ma{B}^2=\frac{g_1^2}{16}, \quad g_c^2=g_s^2\frac{g_1^2}{24},
\label{3.114}
\ee
from which 
\be
{g'}\ma{B}^2=(\frac{1}{g\ma{R}^2}+\frac{1}{g\ma{B}^2})^{-1}=\frac{g_1^2}{40}
\label{3.115}
\ee
follows. Thus, we obtain
\be
\sin^2\theta\ma{W}=\frac38, \qquad \sin^2\theta\ma{Y}=\frac35,
\label{3.116}
\ee
which are values predicted by  $SO(10)$ GUT.
However, in our model the grand unification for coupling constants 
is not achieved because of
the extra parameter $g_s$ in $g_c$ though $g\ma{L}=\sqrt{5/3}g'$.
\par
\subsection{The Higgs boson mass }
The mass spectrum of the Higgs bosons in tree level is also extracted from
the Higgs potential and interacting terms given in Eqs.(\ref{3.66}), 
(\ref{3.79}), (\ref{3.86})$\sim$(\ref{3.88}) and (\ref{3.91}). 
According to Eq.(\ref{3.108}), 
we know the particle contents which are absorbed
by the gauge bosons when the spontaneous breakdown of symmetry takes place.
If $\phi_1^0$ and $\phi_2^0$ are denoted as
\be
   \phi_1^0=\frac{1}{\sqrt{2}}(\phi\ma{1R}^0+i\phi\ma{1I}^0),\qquad
   \phi_2^0=\frac{1}{\sqrt{2}}(\phi\ma{2R}^0+i\phi\ma{2I}^0),
   \label{3.117}
\ee
the following combination of $\phi\ma{1I}^0$ and $\phi\ma{2I}^0$:
\be
   {\phi\ma{I}^{0}}\,' =\phi\ma{1I}^0\cos\beta-\phi\ma{2I}^0\sin\beta
                      \label{3.118}
\ee
is absorbed into the neutral gauge boson, whereas
\be
   \phi\ma{I}^{0} =\phi\ma{1I}^0\sin\beta+\phi\ma{2I}^0\cos\beta
                      \label{3.119}
\ee
remains physical. In these equations, use has been made of
\be
\cos\beta=\frac{\mu_1}{\sqrt{\mu_1^2+\mu_2^2}}=\frac{\mu_1}{2\mu},
\hskip 2cm
\sin\beta=\frac{\mu_2}{\sqrt{\mu_1^2+\mu_2^2}}=\frac{\mu_2}{2\mu}.
\label{3.123}
\ee

Similarly, if the combination of $\phi_1^+$ and $\phi_2^+$ is denoted as
\be
   {\phi^{+}}' =\phi_1^+\cos\beta-\phi_2^+\sin\beta, \label{3.120}
\ee
this field is  absorbed into the charged gauge boson, whereas
\be
   \phi^{+} =\phi_1^+\sin\beta+\phi_2^+\cos\beta  \label{3.121}
\ee
remains physical. After all, five Higgs bosons out of eight components of 
$\phi_1$ and $\phi_2$ become massive. The masses of these Higgs bosons
are obtained below.
Then, the components of $\xi$ absorbed 
into gauge bosons are easily extracted 
from Eq.(\ref{3.45}). $\xi^0$ and $\xi^-$ are absorbed into
$Z\ma{R}$ and $W\ma{R}^-$, respectively. Only surviving Higgs boson is 
the doubly charged component $\xi^{--}$ 
whose mass is at the order of $M\ma{R}$
, as shown below.\par 
For simplicity, $g_1=g_2=g_3=g_4$ is assumed without loss of generality.
In this limit, $\gamma_h^{-1}$ and $\gamma_\xi^{-1}$ are given 
as $g_1^2/24=g^2$ and $g_1^2/6=4g^2$, respectively. 
From Eqs.(\ref{3.66}), 
(\ref{3.79}), (\ref{3.86})$\sim$(\ref{3.88}) and (\ref{3.91}),
we find 
the mass terms of above mentioned surviving Higgs bosons. 
The masses of 
$\phi\ma{I}^0$, $\phi^\pm$, and $\xi^{--}$ are explicitly written as
\ba
&&
m^2_{\phi\ma{I}^0}=M\ma{W_R}^2(\cos^2\beta) +2M_{_W}^2,
\label{3.124}\\
&&
m^2_{\phi^\pm}=M\ma{W_R}^2(\cos\beta+\sin\beta)^2+M_{_W}^2,
\label{3.125}\\
&&
m^2_{\xi^{--}}=4M\ma{W_R}^2+M_{_W}^2\cos^22\beta,
      \label{3.126}
\ea
respectively. However, $\phi\ma{1R}^0$ and $ \phi\ma{2R}^0$ are mixed as 
\ba
M^2_{ \phi\ma{1R}^0,\: \phi\ma{2R}^0}&=&
M_{_W}^2\left\{(\phi\ma{2R}^0\sin\beta+\phi\ma{1R}^0\cos\beta)^2
+2{\phi\ma{2R}^0}^2\sin\beta+2{\phi\ma{1R}^0}^2\cos\beta\right.
\nonum
&&
+\left.\frac{}{}2\mu^2(\phi\ma{2R}^0
\cos\beta+\phi\ma{1R}^0\sin\beta)^2\right\}
+2M\ma{W_R}^2(\phi\ma{2R}^0\cos\beta-\phi\ma{1R}^0\sin\beta)^2.
\label{3.127}
\ea
In the limit of $M\ma{W_R} >\!\!> M_{_W}$, 
we can diagonalize this mass term as
\be
M^2_{ \phi\ma{1R}^0,\: \phi\ma{2R}^0}=
m_{\phi\ma{R}^0}^2{\phi\ma{R}^0}^2
+m_{{\phi\ma{R}^0}'}^2{{\phi\ma{R}^0}'}^2,
\label{3.128}
\ee
where
\be
{\phi\ma{R}^0}={\phi}\ma{2R}^0\sin\beta+{\phi}\ma{1R}^0\cos\beta,
\hskip 1cm
{\phi\ma{R}^0}'={\phi}\ma{2R}^0\cos\beta-{\phi}\ma{1R}^0\sin\beta,
\label{3.129}
\ee
and the masses of ${\phi\ma{R}^0}$ and ${\phi\ma{R}^0}'$ are given as
\ba
   && m_{\phi\ma{R}^0}^2=M_{_W}^2\left(3+\sin^22\beta
                       \right), \label{3.130}\\
  && m_{{\phi\ma{R}^0}'}^2=2M\ma{W_R}^2+
  M_{_W}^2\left(1+\cos^22\beta\right).
                \label{3.131}
\ea
According to the mass values presented above, it is known that
only detectable Higgs boson at the weak energy scale is ${\phi\ma{R}^0}$,
whereas other five Higgs bosons with masses at the scale of $M\ma{R}$
are far beyond the energy scale of the present and near future accelerator.
This is a characteristic feature of the model in the present article.
The approximation $\mu_1\cong \mu_2\cong\mu$ yields  the mass relation
\be
         m_{\phi\ma{R}^0}=2M\ma{W} . \label{3.132}
\ee
Above discussions hold only in tree level. However, we can perform the
renormalization group analyses of these relations under the assumption that
these relations hold at the point that the electromagnetic and weak
interactions are unified as shown in Eqs.(\ref{3.114})$\sim$(\ref{3.116}).
This has been in fact done in the case of the standard model 
in Ref.\cite{RGNCG}.\par
\section{Reconstruction of the fermionic sector in \lr}
The brief summary about the general formulation 
to obtain the Dirac Lagrangian
in NCG is presented with small modification fitting it to the reconstruction
of \lr in this article.
\par
We at first define the covariant derivative 
acting on the spinor $\psi(x,n)$  by
\be
{\cal D}\psi(x,n)=({\bf d}+ A^f(x,n))\psi(x,n), \label{4.1}
\ee
where $A^f(x,n)$
is the differential representation to make ${\cal D}\psi(x,n)$
covariant against the gauge transformation.
In the present formulation stated in Section 3, $A^f(x,n)$ coincides
with ${\cal A}(x,n)$ in Eq.(\ref{2.8}).
Thus, the superscript $f$ in Eq.(\ref{4.1}) is removed, hereafter.
\be
{\cal A}(x,n)
=\left(A_\mu(x,n)+G_\mu(x,n)\right)dx^\mu
+\sum_k {\mit\Phi}_{nk}(x)\chi_k, \label{4.2}
\ee
where it should be noted that the scale transformations of gauge and Higgs 
fields are performed after getting the Dirac Lagrangian.
By setting the algebraic rule 
\ba
d_\chi \psi(x,n)&=&\sum_k d_{\chi_k} \psi(x,n)\nonum
     &=&\sum_k M_{nk}\chi_k\psi(x,n)
     =\sum_k M_{nk}\psi(x,k)\chi_k,
 \label{4.3}
\ea
${\cal D}\psi(x,n)$  is described as
\be
{\cal D}\psi(x,n)
= \left\{\left(\partial_\mu + A_\mu(x,n)+G_\mu(x,n)\right) dx^\mu
+\sum_k H_{nk}(x)\chi_k\right\}\psi(x,n),
\label{4.4}
\ee
with $H_{nk}(x)$ in Eq.(\ref{2.23bc}).\par
\par
Here, we comment on  the parallel transformation 
of the fermion field on the discrete space $M_4\times Z\ma{N}$.
If Eq.(\ref{4.3}) is denoted as
\be
d_\chi \psi(x,n)=\sum_k\p_{nk}\psi(x,k)\chi_k, \label{4.5}
\ee
the covariant derivative ${\cal D}\psi(x,n)$ is rewritten as
\ba
   {\cal D}\psi(x,n)&=&\left\{\frac{}{}(\p_\mu+A_\mu(x,n)
                  +G_\mu(x,n))dx^\mu\psi(x,n)\right. \nonum
 &&\hskip 4cm  \left. +\sum_k(\p_{nk}+{\mit\Phi}_{nk}(x))\psi(x,k)\chi_k
         \right\}.   \label{4.6}
\ea
Equation(\ref{4.6}) implies that ${\mit\Phi}_{nk}(x)\psi(x,k)\chi_k$ 
expresses the variation accompanying the parallel transformation
from $k$-th to $n$-th points on the discrete space just as
$A_\mu(x,n)\psi(x,n)dx^\mu$ is the variation of parallel
transformation on the Minkowski space $M_4$.
This insures that the shifted Higgs field $\mit \Phi_{nk}(x)$ 
is the gauge field on  the discrete space.\par
The nilpotency of $d_\chi$ in this case is also important
to obtain  consistent explanations of covariant derivative and 
parallel transformation. From Eq.(\ref{4.3}), we find
\be
    (d_{\chi_l}d_{\chi_k}+d_{\chi_k}d_{\chi_l})\psi(x,n)=0 \label{4.7}
\ee
which 
 implies that if the Higgs field ${\mit\Phi}_{nk}(x)$
as the gauge field on the discrete space
 vanishes, the curvature on the discrete space also vanishes.
From Eq.(\ref{4.7}), the nilpotency of ${\bf d}$ is evident.
\par
In order to obtain the Dirac Lagrangian, 
we introduce the associated spinor one-form
by
\be
{\tilde {\cal D}}\psi(x,n)= \gamma_\mu \psi(x,n)dx^\mu
              +i\{{\cal O}g\ma{Y}(n)\}^\dagger\psi(x,n)\sum_k\chi_k,
\label{4.8}
\ee
where ${g\ma{Y}}(n)$ is also a $24\times24$ matrix relating to the 
Yukawa coupling constant. 
The operator ${\cal O}$
is defined with an appropriate field $A$ as 
\be
   \left\{{\cal O}g\ma{Y}(n)\right\}A\psi(x,k)
   =A g\ma{Y}(n) \psi(x,k)\label{4.8a}
\ee
if $\psi(x,k)$ is the right-handed fermion field and
\be
   \left\{{\cal O}g\ma{Y}(n)\right\}A= g\ma{Y}(n)A\psi(x,k)  \label{4.8b}
\ee
if $\psi(x,k)$ is the left-handed fermion field. 
More precisely in the later construction, 
Yukawa coupling matrix $g\ma{Y}(n)$ operates directly to 
the right-handed fermion $\psi\ma{R}=\sqrt{2}\psi(x,2)$ in Eq.(\ref{3.1}). 
This prescription 
is needed to insure the Hermiticity of the Dirac Lagrangian.
The Dirac Lagrangian  is obtained by taking the inner product
\ba
{\cal L}\ma{\rm D}(x,n)&=& 
i{\rm Tr}\,<{\tilde {\cal D}}\psi(x,n),{\cal D}\psi(x,n)>\nonum
&=& i\,{\rm Tr}\,
\left[\frac{}{}\,{\bar\psi}(x,n)\gamma^\mu(\partial_\mu
                            +A_\mu(x,n)+G_\mu(x,n))\psi(x,n)
\right.\nonum && \hskip 4cm\left.
+i{\bar\psi}(x,n)({\cal O}g\ma{Y}(n))\sum_k H_{nk}(x)\psi(x,k)\,\right],
\label{4.11}
\ea
where 
we  have used the definitions of the inner products for spinor one-forms
considering  Eq.(\ref{2.34}),
\ba
\a <A(x,n)dx^\mu, B(x,n)dx^\nu>={\rm Tr}\,\bar{A}(x,n)B(x,n)g^{\mu\nu},
\label{4.12}\\
\a <A(x,n)\chi_k, B(x,n)\chi_l>
                  =-\delta_{kl}{\rm Tr}\,\bar{A}(x,n)B(x,n)
                  \label{4.13}
\ea
with other inner products vanishing.
The total Dirac Lagrangian is the sum over $n=1,2\cdots N$
\be
{\cal L}\ma{\rm D}(x)
      = \sum_{n=1}^{N}{\cal L}\ma{\rm D}(x,n). 
\label{4.14}
\ee
\par
Let us move to the explicit reconstruction of the Dirac Lagrangian 
corresponding to the specification of the fermion field $\psi(x,n)$ 
in Eq.(\ref{3.1}). After the rescaling of the gauge and Higgs bosons, 
we can write ${\cal D}\psi(x,n)$ and ${\tilde {\cal D}}\psi(x,n)$
as follows:
\ba
      {\cal D}\psi(x,1)&=& 
      \left[ \p_\mu\otimes1^8 
 -\frac i2 \left\{g\ma{L}\sum_{k=1}^3\tau^k\otimes1^4{A^k\ma{L}}_\mu
   + c g\ma{B}B_\mu
   \right.\right.   \nonum && \left.\left.\hskip 0cm
     +\sum_{a=1}^8 1^2\otimes\lambda'^ag_c G^a_\mu
   \right\}\right]\otimes1^3\psi(x,1)dx^\mu
    +g_h h\otimes 1^4\otimes 1^3 \psi(x,2)\chi_2,
              \label{4.15}\\
              && \frac{}{}\nonum
     {\tilde {\cal D}}\psi(x,1)&=& 
              1^{24}\gamma_\mu \psi(x,1)dx^\mu
              +i\{{\cal O}g\ma{Y}(1)\}^\dagger\psi(x,1)\sum_k\chi_k,
              \label{4.16}\\
      {\cal D}\psi(x,2)&=& \left[ \p_\mu\otimes1^8 
 -\frac i2 \left\{g\ma{R}\sum_{k=1}^3\tau^k\otimes1^4{A^k\ma{R}}_\mu
   + c g\ma{B}B_\mu
   \right.\right.   \nonum && \left.\left.\hskip 4cm
  + \sum_{a=1}^8 1^2\otimes\lambda'^ag_c G^a_\mu
   \right\}\right]\otimes1^3\psi(x,2)dx^\mu
   \nonum  && 
    +g_h h^\dagger\otimes 1^4\otimes 1^3 \psi(x,1)\chi_1
      +g_\xi\xi\ma{R}\otimes b
                    \otimes 1^3 \psi(x,4)\chi_4,
              \label{4.17}\\
              && \frac{}{}\nonum
       {\tilde {\cal D}}\psi(x,2)&=& 
              1^{24}\gamma_\mu \psi(x,2)dx^\mu
              +i\{{\cal O}g\ma{Y}(2)\}^\dagger\psi(x,2)\sum_k\chi_k,
              \label{4.18}\\
      {\cal D}\psi(x,3)&=& \left[ \p_\mu\otimes1^8 
 -\frac i2 \left\{g\ma{L}\sum_{k=1}^3\tau^k\otimes1^4{A^k\ma{L}}_\mu
   - c g\ma{B}B_\mu 
   \right.\right.   \nonum && \left.\left.\hskip 0cm
   -\sum_{a=1}^8 1^2\otimes{\lambda'^a}^\ast g_c G^a_\mu
   \right\}\right]\otimes1^3\psi(x,3)dx^\mu
    +g_h {\tilde h}\otimes 1^4\otimes 1^3 \psi(x,4)\chi_4,
              \label{4.19}\\
              && \frac{}{}\nonum
       {\tilde {\cal D}}\psi(x,3)&=& 
               1^{24}\gamma_\mu \psi(x,3)dx^\mu
              +i\{{\cal O}g\ma{Y}(3)\}^\dagger\psi(x,3)\sum_k\chi_k,
              \label{4.20}\\
      {\cal D}\psi(x,4)&=& \left[ \p_\mu\otimes1^8 
  -\frac i2 \left\{g\ma{R}\sum_{k=1}^3\tau^k\otimes1^4{A^k\ma{R}}_\mu
   - c g\ma{B}B_\mu
   \right.\right.    \nonum && \left.\left.\hskip 4cm
   -\sum_{a=1}^8 1^2\otimes{\lambda'^a}^\ast g_c G^a_\mu
   \right\}\right]\otimes1^3\psi(x,4)dx^\mu
   \nonum  &&  
   +g_h {\tilde h}^\dagger\otimes 1^4\otimes 1^3 \psi(x,3)\chi_3
      +g_\xi\xi\ma{R}^\dagger\otimes b
                    \otimes 1^3 \psi(x,2)\chi_2,
              \label{4.21}\\
              && \frac{}{}\nonum
       {\tilde {\cal D}}\psi(x,4)&=& \frac{}{}
              1^{24}\gamma_\mu \psi(x,4)dx^\mu
              +i\{{\cal O}g\ma{Y}(4)\}^\dagger\psi(x,4)\sum_k\chi_k.
              \label{4.22}
\ea
From these equations together with Eq.(\ref{4.11}),
we can find the Dirac Lagrangian ${\cal L}\ma{\rm D}$
which consists of 
the kinetic term ${\cal L}_{\rm KT}$ 
and the Yukawa interaction term ${\cal L}_{\rm Yukawa}$ between
fermions and Higgs bosons  expressed as    
\ba
 &&{\cal L}_{\rm KT}=
 i{\bar \psi}\ma{L}\gamma^\mu
      \left[ \p_\mu\otimes1^8 
 -\frac i2 \left\{g\ma{L}\sum_{k=1}^3\tau^k\otimes1^4{A^k\ma{L}}_\mu
      \right.\right.   \nonum && \left.\left.\hskip 6cm
   + c g\ma{B}B_\mu
     +g_c\sum_{a=1}^8 1^2\otimes\lambda'^a G^a_\mu
   \right\}\right]\otimes1^3\psi\ma{L}\nonum
 &&\hskip 1cm
+ i{\bar \psi}\ma{R}\gamma^\mu
      \left[ \p_\mu\otimes1^8 
 -\frac i2 \left\{g\ma{R}\sum_{k=1}^3\tau^k\otimes1^4{A^k\ma{R}}_\mu
      \right.\right.   \nonum && \left.\left.\hskip 6cm
   + c g\ma{B}B_\mu
     +g_c\sum_{a=1}^8 1^2\otimes\lambda'^a G^a_\mu
   \right\}\right]\otimes1^3\psi\ma{R}
\label{4.23}\\
 &&{\cal L}_{\rm Yukawa}=
      -{\bar\psi}\ma{L}h\otimes1^4\otimes1^3g\ma{Y}\psi\ma{R}
  -{\bar\psi}\ma{R}h^\dagger g\ma{Y}^\dagger\otimes1^4\otimes1^3\psi\ma{R}
\nonum
&& \hskip 2cm
         -\frac12{\bar\psi}\ma{R}g\ma{Y}^\dagger\xi\ma{R}\otimes b
         \otimes 1^3\psi\ma{R}^c
         -\frac12{\bar\psi}\ma{R}^c \xi\ma{R}^\dagger\otimes b
         \otimes 1^3g\ma{Y}\psi\ma{R},
  \label{4.24}
\ea
where
$\psi\ma{L}=\sqrt{2}\psi(x,1)$ and $\psi\ma{R}=\sqrt{2}\psi(x,2)$
by use of Eq.(\ref{3.1}). In introducing ${\cal L}_{\rm Yukawa}$,
$g\ma{Y}(1)=g\ma{Y}$, $g\ma{Y}(2)=g\ma{Y}^\dagger$, 
$g\ma{Y}(3)=g\ma{Y}^\dagger$ and $g\ma{Y}(4)=g\ma{Y}^T$ are assumed
to insure the Hermiticity of the Yukawa interaction where $g\ma{Y}$
is the Yukawa coupling constant written in $24\times 24$ matrix as
\be
    g\ma{Y}={\rm diag}(g^u,g^u,g^u,g^\nu,g^d,g^d,g^d,g^e).
                               \label{4.25}
\ee
$g^u$, $g^d$, $g^\nu$ and $g^e$ in Eq.(\ref{4.25}) are 
complex Yukawa coupling constants denoted in
$3\times3$ matrix in generation space.
\par
We address 
the $U(1)\ma{Y}$ hypercharge matrix 
of left and right handed leptons and quarks.
Owing to the vacuum expectation value of $\xi\ma{R}$, 
$SU(2)\ma{R}\times U(1)$ gauge symmetry breaks down to $U(1)\ma{Y}$ gauge
symmetry. The resulting $U(1)$ gauge boson is $B'_\mu$ in Eq.(\ref{3.106}).
Inserting the following equations
\ba
&&  B_\mu=B'_\mu\cos\theta\ma{Y}
     -{Z\ma{R}}_\mu\sin\theta\ma{Y},\label{4.26}\\
&&  {A\ma{R}^3}_\mu=B'_\mu\sin\theta\ma{Y}
   +{Z\ma{R}}_\mu\cos\theta\ma{Y}\label{4.27}
\ea
into Eq.({\ref{4.23}) and using Eqs.(\ref{3.96}) and (\ref{3.100}),
it is known that
the $U(1)\ma{Y}$ hypercharge matrix 
of $\psi\ma{L}$ is $c$ itself and that of 
$\psi\ma{R}$ is 
\be
c+\tau^3\otimes 1^4={\rm diag}\left(\frac43,\;\frac43,\;\frac43,0,\;
-\frac13,\;-\frac13,\;-\frac13,\;-2\right), \label{4.28}
\ee
which correctly reproduce the $U(1)\ma{Y}$ 
hypercharge of $\psi\ma{L}$ and
$\psi\ma{R}$, respectively.\par
It should be remarked that leptons does not interact with the
color gauge boson owing to ${\lambda'}^a$ in Eq.(\ref{3.5a}).
 The Yukawa interaction of lepton is expressed as
\ba
 &&{\cal L}_{\rm Yukawa}^l=
      -{\bar l}\ma{L}h\otimes1^3g\ma{Y}^l l\ma{R}
  -{\bar l}\ma{R}h^\dagger {g\ma{Y}^l}^\dagger\otimes1^3 l\ma{R}
\nonum
&& \hskip 2cm
         -\frac12{\bar l}\ma{R}{g\ma{Y}^l}^\dagger\xi\ma{R}
         \otimes 1^3 l\ma{R}^c
         -\frac12{\bar l}\ma{R}^c \xi\ma{R}^\dagger 
         \otimes 1^3g\ma{Y}^l l\ma{R},
  \label{4.29}
\ea
where $l\ma{L}=(\nu\ma{L},\;e\ma{L})^t$,
$l\ma{R}=(\nu\ma{R},\;e\ma{R})^t$ and
$l\ma{R}^c=(e\ma{R}^c,\;-\nu\ma{R}^c)^t$. $g\ma{Y}^l$ is a lepton part
of $g\ma{Y}$ in Eq.(\ref{4.25}).
If the mass term is extracted  from Eq.(\ref{4.29}), 
it is evident that the seesaw mechanism for the left and right handed
neutrino \cite{Gell} works well.\par
Quarks do not interact with $\xi$ owing to the matrix $b$ 
in Eq.(\ref{3.13}). The Yukawa interaction of quarks is written as 
\be
 {\cal L}_{\rm Yukawa}^q=
      -{\bar q}\ma{L}h\otimes1^9g\ma{Y}^q q\ma{R}
  -{\bar q}\ma{R}{g\ma{Y}^q}^\dagger h^\dagger \otimes1^9 q\ma{L},
  \label{4.30}
\ee
where $q\ma{L}=(u\ma{L},\;d\ma{L})^t$ and
$q\ma{R}=(u\ma{R},\;d\ma{R})^t$. $g\ma{Y}^q$ is a quark part of $g\ma{Y}$.
Up and down quarks  obtain the masses
through the vacuum expectation value of $h$ and 
the Kobayashi-Maskawa matrix comes out by diagonalizing the generation
mixing matrices $g^u$ and $g^d$ in $g\ma{Y}$ in Eq.(\ref{4.25}).
\section{Concluding remarks}
\par
We have reconstructed the left-right symmetric gauge model based on the
new scheme of our noncommutative differential geometry on the discrete space
$M_4\times Z_4$ which is a product space of the Minkowski space multiplied 
by four point space. The Higgs field is well explained as a part of
the generalized connection of the principal bundle with the base space 
$M_4\times Z\ma{N}$, as shown in Eq.(\ref{2.8a}). One-form basis $\chi_k$
on the discrete space is the fundamental 
ingredient in the present formulation 
to hold the noncommutative nature that $\chi_n\wedge\chi_k$ is independent
of $-\chi_k\wedge\chi_n$.  
In general, though the NCG approach can propose the Higgs potential 
which automatically yields the spontaneous breakdown of gauge symmetry,
it imposes the severe constraints on Higgs potential and interacting terms.
In Ref.\cite{WSLR}, we could not obtain the Higgs potential terms 
sufficient to explain the phenomenology of \lr. Thus, we revised an algebraic
rule so as to relax the severe constraint and meet the requirements of
the phenomenology. This modification is shown in Eq.(\ref{2.14c}). 
In addition, we adopt the fermion field to be a vector in a 24-dimensional
space including the generation space together with 
the space of flavor and color symmetry. This fermion field has the same
placement of lepton and quark as in SO(10) GUT. As the Higgs bosons, 
 we incorporate an adjoint $\xi\ma{R}$ and $h$ 
 containing two Higgs doublet bosons, which are transformed as 
 $(1,3,-2)$ and $(2,2^\ast,0)$ under $SU(2)\ma{L}\times SU(2)\ma{R}\times
 U(1)$ symmetry, respectively. $h$ and $\xi\ma{R}$ are the
 standard ingredients in \lr. In addition to $h$, ${\tilde h}$ is also 
 considered. Owing to these considerations, we could extend the 
 discussions about the gauge and Higgs boson masses in section 3.
 In the case of $g_1=g_2=g_3=g_4$, we reach the interesting relations
 $\sin^2\theta\ma{W}=3/8$ and $\sin^2\theta\ma{Y}=3/5$ which are 
 equal to the predictions of GUT. However, the grand unification 
 of coupling constants is not achieved in this formulation. 
Judging from this, the case of $g_1=g_2=g_3=g_4$ indicates that 
the electromagnetic and weak interactions are unified at this point.
As for the Higgs bosons, the surviving boson 
at weak energy scale is only one Higgs component, others have
 masses at energy scale of $M\ma{R}$. This is very characteristic
 feature in the present formulation.
 The mass of this surviving boson is $2M\ma{W}$ at 
 the case of $g_1=g_2=g_3=g_4$. It is very interesting to make 
 the renormalization group analysis of this mass relation down to the
 weak energy scale. This type of analysis has already been 
 done in \cite{RGNCG}. According to this analysis, the Higgs boson mass
 is 164GeV if the top quark mass is 175GeV.
 The similar analysis will be done in near future.

\vskip 0.5cm
\par
\section*{\bf Acknowledgements}
The author would like to
express his sincere thanks to
Professors J.~Iizuka,
 H.~Kase, K.~Morita and M.~Tanaka 
for useful suggestion and
invaluable discussions on the non-commutative geometry.
\vskip 0.5cm
\def\jmp{J.~Math.~Phys.$\,$}
\def\pl{Phys. Lett.$\,$ }
\def\np{Nucl. Phys.$\,$}
\def\ptp{Prog. Theor. Phys.$\,$}
\def\prl{Phys. Rev. Lett.$\,$}
\def\pr{Phys. Rev. D$\,$}
\def\mp{Int. Journ. Mod. Phys.$\,$ }


\begin{thebibliography}{99}
\bibitem{Con}
A.~Connes, p.9 in {\it The Interface of Mathematics and Particle
Physics}, \hfill\break
ed. D.~G.~Quillen, G.~B.~Segal, and Tsou.~S.~T.,
Clarendon Press, Oxford, 1990. See also,
Alain Connes and J. Lott, 
Nucl. Phys. {\bf B}(Proc. Suppl.) {\bf 18B} (1990) 57.
\bibitem{Cham} A.~H.~Chamseddine, G.~Felder and J.~Fr\"olich,
\pl {\bf B296}, 109(1992); \ \np B395 (1993) 672;\ 
A.~H.~Chamseddine and J.~Fr\"olich,
\pr {\bf 50} (1994) 2893.
\bibitem{Kast}
D.~Kastler, Rev. Math. Phys. {\bf 5}(1993) 477.
\bibitem{Dubo}
M.~Dubois-Violette, Class. Quantum. Grav. {\bf 6}(1989) 1709; 
M.~Dubois-Violette, R.~Kerner, and J.~Madore, J. Math.
Phys. {\bf 31} (1990)  316.
\bibitem{Coqu}
R.~Coquereaux, G.~Esposito-Farese, and G.~Vaillant, \np  {\bf B353}
(1991) 689;\   
R.~Coquereaux, G.~Esposito-Farese and 
  F.~Scheck, \mp {\bf A7} (1992) 6555; R.~Coquereaux, R.~Haussling,
  N.~Papadopoulos and F.~Scheck, {\it ibit}. {\bf 7} (1992) 2809.
\bibitem{Bala}
B.~Balakrishna, F.~G{\"u}rsey and K.C.~Wali, \pl {\bf B254} (1991) 430;
  \pr {\bf 46} (1992) 6498.    
\bibitem{Sita} 
A.~Sitarz, \pl,\ {\bf B308} (1993) 311, Jour. Geom. Phys. {\bf 15}(1995) 123.
\bibitem{Gou}
H-G.~Ding, H-Y.~Gou, J-M.~Li and K.~Wu, 
 Zeitschift f\"ur Physik, 
       {\bf C64}(1994)  521.
\bibitem{Naka}
S.~Naka and E.~Umezawa, \ptp {\bf 92} (1994) 189.
\bibitem{WS} K.~Morita and Y.~Okumura, \ptp {\bf 91} (1994) 959.
\bibitem{SU5}
Y.~Okumura, \pr {\bf 50} (1994) 1026.
\bibitem{SO10} 
Y.~Okumura, \ptp, {\bf 94}  (1995) 607.
\bibitem{FSTM}
Y.~Okumura,
\ptp {\bf 95} (1996) 969.
\bibitem{WSLR}
Y.~Okumura and K.~Morita,  
Nuovo Cimento {\bf 109A} (1996) 311.
\bibitem{BRST}
Y.~Okumura, Phys. Rev. {\bf D54} (1996) 4114. 
\bibitem{NCGNP}
Y.~Okumura, preprint"
 Non-Commutative Differential Geometry on Discrete Space $M_4\times Z_{_N}$
and Gauge Theory" to appear in \ptp, {\bf 96} (1996).
\bibitem{RGNCG}
Y.~Okumura, preprint CHUBU9608, 
"Renormalization group analysis of the Higgs boson mass 
in a noncommutative differential geometry", hep-ph/9608208.
\bibitem{Gell}
M.~Gell-Mann, P.~Ramond and R.~Slanski, in {\it Supergravity}, 
Proceeding of the Workshop, Stony Brook, New-York, 1979, edited 
by P. van Nieuwenhuizen and D.~Freedman(North-Holland, Amsterdam, 1980); 
T.~Yanagida, in {\it Proceeding of the Workshop on Unified Theories 
and Baryon Number in the Universe}, Tsukuba, Japan, 1979,edited 
by A.~Sawada and A.~Sugamoto(KEK Report No.79-18 Tsukuba, 1979);
R.~N.~Mohapartra and G.~Senjanovic, Phys. Rev. Lett. {\bf 44} (1980) 912 
and the
references in
M.~Fukugita and T.~Yanagida, 
Physics of Neutrinos in {\it Physics and Astrophysics of Neutrinos}, 
edited by M.~Fukugita and A.~Suzuki, 
Springer-Verlag, Tokyo, 1994. 
\end{thebibliography}
\end{document}